\documentclass[12pt,letterpaper]{article}
\usepackage{amsmath,amssymb,array,calc,rotating,epsfig,psfrag,amscd, cite}

\newcommand{\nc}{\newcommand}

 \nc{\ra}{\rightarrow}
\def\al{\alpha}

\nc{\veps}{\varepsilon}
\def\gam{\gamma}

\def\om{\omega}
\nc{\vphi}{\varphi}
\def\tha{\theta}

\def\Gam{\Gamma}

\def\Om{\Omega}
\def\Sig{\Sigma}

\nc{\bea}{\begin{eqnarray}}
\nc{\eea}{\end{eqnarray}}
\nc{\be}{\begin{equation}}
\nc{\ee}{\end{equation}}
\nc{\cA}{{\cal A}}
\nc{\cB}{ \cal B}

\nc{\cF}{{\cal F}}
\nc{\cG}{{\cal G}}

\nc{\cL}{{\cal L}}
\nc{\M}{{\cal M}}
\nc{\cM}{{\cal M}}
\def\N{{\cal N}}

\nc{\cQ}{{\cal Q}}
\nc{\cR}{{\cal R}}

\def\T{{\cal T}}

\def\W{{\cal W}}

\nc{\BB}{{\mathbb B}}
\nc{\CC}{{\mathbb C}}
\nc{\DD}{{\mathbb D}}
\nc{\EE}{{\mathbb E}}
\nc{\FF}{{\mathbb F}}
\nc{\GG}{{\mathbb G}}
\nc{\HH}{{\mathbb H}}
\nc{\JJ}{{\mathbb J}}
\nc{\RR}{{\mathbb R}}
\nc{\PP}{{\mathbb P}}
\nc{\QQ}{{\mathbb Q}}
\nc{\ZZ}{{\mathbb Z}}
\nc{\CP}{{\CC\PP}}
\nc{\calone}{{\mathbb 1}}
\nc{\half}{\frac{1}{2}}
\nc{\qrt}{\frac{1}{4}}
\nc{\del}{\partial}

\nc{\delbar}{\bar\partial}
\nc{\Spin}{\operatorname{Spin}}
\nc{\SO}{\operatorname{SO}}

\nc{\Sp}{{\rm Sp}}
\nc{\com}[2]{{ \left[ #1, #2 \right] }}
\nc{\acom}[2]{{ \left\{ #1, #2 \right\} }}
\nc{\rr}{\rightarrow}
\nc{\p}{\partial}
\nc{\LT}{{\LL_\T}}
\nc{\Tr}{{\rm Tr}}
\nc{\tr}{{\rm tr}}
\def\com#1#2{{ \left[ #1, #2 \right] }}
\def\acom#1#2{{ \left\{ #1, #2 \right\} }}
\nc{\tKT}{\widetilde{K3}}
\nc{\ttha}{\tilde{\theta}}
\nc{\tphi}{\tilde{\phi}}
\nc{\tPhi}{\tilde{\Phi}}
\nc{\tpsi}{\tilde{\psi}}
\nc{\tgam}{\tilde{\gam}}
\nc{\tGam}{\tilde{\Gam}}
\nc{\tSig}{\tilde{\Sig}}
\nc{\tc}{\tilde c}
\nc{\te}{\tilde e}
\nc{\tg}{\tilde g}
\nc{\tj}{\tilde j}
\nc{\tp}{\widetilde{p}}
\nc{\tq}{\widetilde{q}}
\nc{\ts}{{\tilde s}}
\nc{\tz}{\tilde z}
\nc{\tD}{{\tilde D}}
\nc{\tE}{{\tilde E}}
\nc{\tG}{{\tilde G}}
\nc{\tH}{{\tilde H}}
\nc{\tM}{{\tilde M}}
\nc{\tN}{{\tilde N}}
\nc{\tP}{{\tilde P}}
\nc{\tQ}{{\tilde Q}}
\nc{\tS}{\tilde{S}}
\nc{\tF}{\tilde{{\cal F}}}
\nc{\tX}{\widetilde{X}}
\nc{\hb}{\hat b}
\nc{\hc}{\hat c}
\nc{\hd}{\hat d}
\nc{\he}{\hat e}
\nc{\hf}{\hat f}
\nc{\hg}{\hat g}
\nc{\hh}{\hat h}
\nc{\hp}{\hat p}
\nc{\hw}{\hat w}
\nc{\hx}{\hat x}
\nc{\hy}{\hat y}
\nc{\hz}{\hat z}
\nc{\hA}{\widehat{A}}
\nc{\hE}{\widehat{E}}
\nc{\hH}{\widehat{H}}
\nc{\hJ}{\widehat{J}}
\nc{\tK}{\widetilde{K}}
\nc{\hM}{\widehat M}
\nc{\hF}{\widehat{\F}}

\nc{\ha}{\widehat \alpha}
\nc{\hphi}{\hat{\phi}}
\nc{\hpsi}{\hat{\psi}}
\nc{\hgam}{\hat{\gam}}
\nc{\hPhi}{\hat{\Phi}}
\nc{\hPsi}{\hat{\Psi}}
\nc{\hGam}{\hat{\Gam}}

\nc{\w}{\wedge}

\nc{\ol}{\overline}
\nc{\abar}{\ol{a}}
\nc{\bbar}{\ol{b}}
\nc{\cbar}{\ol{c}}
\nc{\ebar}{\ol{e}}
\nc{\ibar}{\ol{\imath}}
\nc{\jbar}{\ol{\jmath}}
\nc{\kbar}{\ol{k}}
\nc{\lbar}{\ol{l}}
\nc{\mbar}{\ol{m}}
\nc{\nbar}{\ol{n}}
\nc{\ubar}{\ol{u}}
\nc{\vbar}{\ol{v}}
\nc{\wbar}{\ol{w}}
\nc{\xbar}{\ol{x}}
\nc{\ybar}{\ol{y}}
\nc{\zbar}{\ol{z}}

\nc{\Ebar}{\ol{E}}
\nc{\Jbar}{\ol{J}}
\nc{\Qbar}{\ol{Q}}
\nc{\Wbar}{\ol{W}}
\nc{\Xbar}{{\overline X}}
\nc{\Ybar}{{\overline Y}}
\nc{\Zbar}{{\overline Z}}

\nc{\epsbar}{\ol{\epsilon}}
\nc{\lambar}{\ol{\lambda}}
\nc{\psibar}{\ol{\psi}}
\nc{\Psibar}{\ol{\Psi}}
\nc{\phibar}{\ol{\phi}}
\nc{\Phibar}{\ol{\Phi}}
\nc{\chibar}{\ol{\chi}}
\nc{\ombar}{\ol{\om}}
\nc{\Ombar}{\ol{\Om}}
\nc{\bah}{{\mathbf {\hat{A}}}}
\nc{\bX}{{\mathbf X}}
\nc{\dal}{\dot{\al}}
\nc{\thab}{\bar{\theta}}
\nc{\thal}{\theta^{\al}}
\nc{\thdal}{\bar{\theta}^{\dal}}

\nc{\thsigthm}{\tha \sigma^m \thab}
\nc{\thsigthn}{\tha \sigma^n \thab}

\nc{\Dal}{D_{\al}}
\nc{\Ddal}{\bar{D}_{\dal}}
\nc{\CDal}{{\cal D}_{\al}}
\nc{\CDdal}{\bar{\cal D}_{\dal}}

\nc{\eq}[1]{(\ref{#1})}
\nc{\non}{\nonumber}
\nc{\comment}[1]{{\bf #1}}
\nc{\xs}{\not\!\!X}
\nc{\ps}{\not\!\!P}
\nc{\dif}{{d}}
\nc{\equ}{{\rm eq}}

\nc{\AdS}{{\rm AdS}}
\nc{\vol}{{\rm vol}}
\nc{\Ainf}{A_{\infty}}
\nc{\End}{{\rm End}}
\nc{\Ext}{{\rm Ext}}
\nc{\Hom}{{\rm Hom}}
\nc{\IIB}{{\rm IIB}}
\nc{\Pic}{{\rm Pic}}
\nc{\bra}[1]{\langle{#1}|}
\nc{\ket}[1]{|{#1}\rangle}
\nc{\braket}[2]{\langle{#1}|{#2}\rangle}
\nc{\sect}[1]{Section~\ref{#1}}
\nc{\fig}[1]{Fig.~\ref{#1}}
\nc{\chap}[1]{Chapter~\ref{#1}}
\nc{\Dslash}{\ensuremath \raisebox{0.025cm}{\slash}\hspace{-0.32cm} D}
\nc{\no}{\!:\!\!}
\nc{\bpm}{\begin{pmatrix}}
\nc{\epm}{\end{pmatrix}}
 \nc{\bitem}{\begin{itemize}}
 \nc{\eitem}{\end{itemize}}

\newcommand{\C}[1]{$(\ref{#1})$}

\def\Z{\mathbb{Z}}

\def\SO{\operatorname{SO}}

\nc{\rank}{{\rm rank}}
\nc{\pr}{{\rm pr}}

\nc{\tom}{\tilde{\om}}
\nc{\tOm}{\tilde{\Om}}

\def\iibdil{\varphi_{B}}
\def\hetdil{\varphi_{het}}
\def\typeidil{\varphi_{I}}
\def\bwone{B_{w_1}}
\def\bwtwo{B_{w_2}}

\def\btwoi{B_{ y_i w_2}}
\def\btwoj{B_{y_j w_2}}
\def\bonei{B_{y_i w_1}}
\def\bonej{B_{y_j w_1}}

\begin{document}

\begin{titlepage}

\begin{center}

{March 22, 2009}\hfill     EFI-08-12, \, MIFP-08-08 \, NSF-KITP-09-24

\vskip 2 cm
{\Large \bf  Torsional Heterotic Geometries}\\
\vskip 1.25 cm
Katrin Becker\footnote{kbecker@physics.tamu.edu} and Savdeep Sethi\footnote{sethi@uchicago.edu} \\

{ \vskip 1cm $^1${\it Department of Physics, Texas A\&M University,\\}{ \it College Station, TX 77843, USA}\\}

\vskip .2 cm

{ \vskip 0.5cm $^2${\it Enrico Fermi Institute, University of Chicago,\\}{ \it Chicago, IL 60637, USA}\\}

\end{center}

\vskip 0.2 cm

\begin{abstract}
\baselineskip=18pt

We construct new examples of torsional heterotic backgrounds using duality with orientifold flux compactifications.  We explain how
duality provides a perturbative solution to the type I/heterotic string Bianchi identity.
The choice of connection used in the Bianchi identity plays an important role in the construction.
We propose the existence of a much larger landscape of compact torsional geometries using string duality.
Finally, we present some quantum exact metrics that correspond to NS5-branes placed on an elliptic space. These metrics describe how torus isometries are broken by NS flux.


\end{abstract}

\end{titlepage}


\section{Introduction}
\label{introduction}
\baselineskip=18pt

Generic string compactifications involve both non-trivial metrics and non-trivial background fluxes. Most of the work devoted to flux compactifications has been in the context of type II string theory. The analysis of such backgrounds is typically restricted to the supergravity approximation because of the difficulties with quantizing strings in RR backgrounds.
Yet from the original construction of type IIB flux backgrounds, it has been clear that all the interesting physics that arises from type II fluxes must also be found in generic heterotic string compactifications. Such compactifications only involve NS fields: namely, the metric and the three-form flux ${\cal H}_3$. Compactifications with non-trivial ${\cal H}_3$ are known as torsional compactifications~\cite{Strominger:1986uh, Hull:1986kz}.

For example, the large degeneracy of type IIB solutions parametrized by the choice of RR and NS three-form fluxes can, in specific examples, be mapped directly to a large choice of metrics and ${\cal H}_3$ fluxes in the heterotic string. The resulting metrics differ by what has become known as the choice of ``geometric flux.'' This is a special feature of the examples constructed in~\cite{Dasgupta:1999ss}\ which involve torus factors; for a review, see~\cite{Wecht:2007wu}.

Torsional heterotic backgrounds stand a much better chance of admitting tractable world-sheet descriptions than their type II counterparts. Indeed, there is unlikely to be any perturbative string description of a generic type II flux compactification because the string coupling is typically not a free parameter. Rather the control parameter is the volume of the compactification which is why supergravity can be employed. On the other hand in heterotic flux compactifications,  the string coupling is a modulus so conformal field theory can be used to access small volume physics.

The heterotic string is also a natural setting for building realistic models of particle physics. The typical approach taken in the past has been to specify a compact K\"ahler six-dimensional space with metric satisfying
\be
R_{\mu\nu} = 0
\ee
together with a holomorphic gauge bundle. This data can be used as the starting point for defining a $(0,2)$ worldsheet superconformal field theory. For appropriate choices of gauge bundle, it is easy to generate space-time GUT groups like $E_6, SO(10)$ and $SU(5)$. In recent years, there has been striking progress in the construction of physically interesting Calabi-Yau compactifications with bundles that give rise
to a particle content very close to the Standard Model; for some recent references, see~\cite{Anderson:2007nc, Andreas:2007ei, Bouchard:2005ag, Buchmuller:2005jr, Braun:2005nv}.

Nevertheless, these compactifications all suffer from moduli problems. Ideally, we would like to be able to generalize these phenomenologically interesting constructions to torsional backgrounds where most of the moduli can be fixed. The one problematic modulus is the string coupling itself whose stabilization requires non-perturbative physics.

For these collective reasons, we would like to understand the physics of the heterotic string with NS flux better. One of the impediments to progress in this area has been the lack of four-dimensional compact examples.  We do not yet have torsional analogues of the algebro-geometric constructions available for Calabi-Yau spaces. The primary examples of torsional backgrounds
are DRS geometries which involve torus fibrations over $K3$ surfaces and a varying dilaton~\cite{Dasgupta:1999ss}. If the $K3$ surface is related to a quotient of a torus, these geometries are related to quotients of nilmanifolds. These geometries were shown to admit no K\"ahler metrics in~\cite{Goldstein:2002pg}. There have been interesting recent constructions of torsional heterotic backgrounds using nilmanifolds but with a constant dilaton appearing in~\cite{Fernandez:2008wa}.

The first goal of this analysis is to construct new classes of torsional heterotic compactifications. We will do this by generalizing the original DRS construction which started from $K3 \times T^2$ to more general orientifold three-folds. This will give us a family of new torsional solutions.  Along the way, we will explain how the duality map provides a perturbative solution to the heterotic Bianchi identity
\be \label{bianchi}
d {\cal H}_3 = { \alpha' \over 4}  \left[\tr (R \wedge R) - \tr (F \wedge F)  \right]
\ee
where ${\cal H}_3$ satisfies
\be
 {1\over 2\pi \alpha'} \int {\cal H}_3 \in 2 \pi \Z.
\ee
This point was not made explicit in the original construction but has become interesting in light of the work of~\cite{Fu:2006vj}. We will see that the choice of connection used to compute curvatures is central, and a particular connection is preferred in order to retain both simple equations of motion and a simple form for the spinor supersymmetry variations including assorted $\alpha'$ corrections.

That choice corresponds to the curvature two-form $R$ computed with a particular ${\cal H}$-connection denoted $\Omega_+$.\footnote{The choice of $\Omega_+$ versus $\Omega_-$ in the Bianchi identity is only meaningful relative to the sign of ${\cal H}$ appearing in the gravitino variation given in equation~\C{sugravariations}. We can always send ${\cal B}_2 \rightarrow - {\cal B}_2$ to flip the conventions.}
For other choices of connection like the Hermitian connection~\cite{Strominger:1986uh}, the Bianchi identity gives a complex equation of Monge-Amp\`ere type while for the preferred choice, the Bianchi identity is related by duality to an equation of Laplace type where the existence of solutions is immediate.

We will then propose a generalization of this construction whose form is suggested heterotic/F-theory duality. This provides, in part, an explanation for the role played by $G$-flux in modifying heterotic/F-theory duality though there is clearly much more to be understood. For recent progress on understanding the heterotic dual of $G$-flux localized on the discriminant locus of F-theory compactifications (which becomes part of the data specifying the heterotic bundle), see~\cite{Hayashi:2008ba, Donagi:2008ca}\, which extends earlier work~\cite{Curio:1998bva}.

Our proposed form for the metrics and torsion given in
section~\ref{method}\ uses a semi-flat approximation to a smooth
elliptic metric. This leads to a well defined problem of proving the
existence of smooth metrics which agree with this structure except
near singularities of the elliptic fiber. These metrics should solve
the heterotic equations of motion up to $O(\alpha'^2)$. At
$O(\alpha'^3)$, there are new corrections to the equations of motion
from $R^4$ type couplings. They should also satisfy the supergravity
spinor variations which first receive corrections at $O(\alpha'^2)$
in the preferred basis of fields where the Bianchi identity involves
the $\Omega_+$ connection. Perhaps an existence theorem can be
proven by extending the analysis of~\cite{Fu:2006vj, Becker:2006et}.

In section~\ref{checkingSUSY},  we examine the whether our proposed
metrics and fluxes satisfy the local supersymmetry conditions. We
then generalize the metric ansatz to include a torus fiber with
varying volume. These metrics depend on two holomorphic parameters
rather than just the complex structure of the torus, and describe
non-geometric heterotic compactifications. We describe a class of
such solutions.

In the final section, we turn to the question of quantum corrections in the presence of NS ${\cal H}_3$-flux. While little can be said about exact metrics in complex dimension three, there is an exact metric describing NS5-branes on an elliptic space that captures all the quantum corrections which break both isometries of the elliptic fiber. This is roughly a torsional version of the metric found in~\cite{Ooguri:1996me}. It should play an interesting role in repairing singularities of these torsional backgrounds.

\vskip 0.5cm
\noindent
{\bf Note added:} During the completion of this project, we received a paper with additional compact torsional heterotic solutions constructed as orbifolds~\cite{Becker:2008rc}, and some papers with related observations~\cite{Evslin:2008zm, Andriot:2009fp}.

\section{Setting the Stage}
\label{stage}

\subsection{Some background}

We would like to find data to define heterotic $(0,2)$ world-sheet
sigma models. This data involves a metric and a bundle together with
a choice of ${\cal H}_3$-flux. As a matter of notation, we will use ${\cal H}$ or
${\cal H}_3$ to denote the heterotic NS flux and $H$ or $H_3$ to denote the
type II NS flux. The associated gauge potentials are denoted ${\cal
B}_2$ and $B_2$, respectively. The standard notation $F_n$ will be used for the RR
fluxes of type II string theory defined in~\C{deffieldstrengths}\
with associated potentials $C_{n-1}$. For type I string theory,  we
use the notation $F_n'$ for the RR fluxes.

Compactifications with ${\cal H}_3$ are typically stringy because
they involve cycles with size of order $\alpha'$. However, there is
a reasonable but unproven belief that background data satisfying the
heterotic supergravity equations of motion supplemented with the
Bianchi identity~\C{bianchi}\ will suffice to define perturbatively
conformal sigma models; see, in particular~\cite{Sen:1986mg,
Hull:1986kz}.

These conditions are definitely not sufficient to guarantee
non-perturbative conformal invariance even for K\"ahler
compactifications with large volume limits~\cite{Dine:1986zy}.
However, there are special cases like models built from linear sigma
models which can be shown to be non-perturbatively
conformal~\cite{Basu:2003bq, Beasley:2003fx}. It would be ideal to
find analogous constructions for torsional compactifications where
the problem is more acute because of the lack of a large volume
limit; for some steps in this direction, see~\cite{Adams:2006kb, Adams:2009av}. In
this work, however, our goal will be to find more solutions of the
type I/heterotic supergravity equations which can be used as
starting points for a world-sheet analysis.

Now the $Spin(32)/\Z_2$ heterotic string is equivalent to the type I
string via a strong-weak coupling duality. What duality naturally
provides for us are new type I torsional solutions. The relation
between the type I and heterotic solutions is unambiguous at the
level of supergravity but might differ at higher orders in the
$\alpha'$ expansion by field redefinitions.

To understand the conditions of space-time supersymmetry and which choices of connection are permissible, it is
simplest to start with the heterotic space-time effective action
\begin{equation}
\begin{split}
S={1\over 2\kappa^2} \int d^{10}x \sqrt{-g} \, e^{-2 \hetdil}& \Big[R
+4 (\partial \hetdil)^2-{1\over 2} \mid {\cal H} \mid^2 \cr & -
{\alpha' \over 4} \left( \tr \mid {\cal F}\mid^2 -\tr \mid
R_+\mid^2\right)   +O(\alpha'^2) \Big],
\end{split}
\end{equation}
where
\begin{equation}
\tr \mid R_+\mid^2={1\over 2} R_{MNAB} (\Omega_+)R^{MNAB}
(\Omega_+)
\end{equation}
and $ {\cal F}$ is the Yang-Mills field strength.
The Einstein-Hilbert term is constructed using the standard metric
connection denoted $\Omega$, while the Riemann tensor appearing in the
$O(\alpha')$ correction is constructed using the connection $\Omega_+$
where
\begin{equation}\label{conn}
{\Omega^{AB}_\pm}_M= {\Omega^{AB}}_M\pm \half {{\cal
H}^{AB}}_M+O(\alpha'),
\end{equation}
and ${\Omega^{AB}}$ is the spin connection. The definition of ${\cal H}$ already
includes $O(\alpha')$ corrections,
\begin{equation}\label{aaav}
{\cal H} = d {\cal B}_2 +\frac{\alpha'}{4}   \left[ {\rm CS}(\Omega_+) - {\rm
CS}(A)  \right] ,
\end{equation}
where $A$ is the connection on the gauge-bundle. For a summary
of notation, see Appendix~\ref{notation}.

The supersymmetrization of
the four derivative interactions including $R^2$ and the Lorentz
Chern-Simons couplings have been worked out with various choices of
fields in~\cite{Chemissany:2007he, Bergshoeff:1989de,
Bergshoeff:1988nn, Metsaev:1987zx}. The equations of motion arising
from this action are
\bea \label{eom}
 R-4 (\nabla \hetdil)^2+4 \nabla^2 \hetdil -{1\over 2 } \mid {\cal
H}\mid^2-{\alpha'\over 4} \left( \tr \mid {\cal F} \mid^2 - \tr \mid
R_+\mid^2\right) &=& O(\alpha'^2), \cr
 R_{MN}+2 \nabla_M \nabla_N
\hetdil -{1\over 4} {\cal H}_{MAB} {{\cal H}_N}^{AB}
-{\alpha'\over 4} \Big[\tr F_{MP}{F_N}^P && \\  -R_{MPAB}(\Omega_+)
R_{N}^{~~PAB}(\Omega_+)\Big] &=& O(\alpha'^2), \cr
 d \left( e^{-2
\hetdil} \star {\cal H}\right) &=& O(\alpha'^2), \cr
 e^{2\hetdil} d (e^{-2 \hetdil} \star {\cal F})+ {\cal A} \wedge \star {\cal F} -
\star {\cal F} \wedge {\cal A} + {\cal F} \wedge \star {\cal
H}& =& O(\alpha'^2). \non
\eea
The dilaton equation of motion has been used to simplify the
Einstein equation appearing above. In order to obtain these equations,
it is easiest to compute the variation of the action with respect to the fields
$\hetdil$, $g_{MN}$, $B_{MN}$, $A_M$ appearing explicitly and then
the variation with respect the the connection $\Omega_+$ which
implicitly also depends on these variables. According to a
lemma proven in~\cite{Bergshoeff:1989de}, the variation of the
$\alpha'$ correction to the action with respect to $\Omega_+$ is
proportional to the leading order equations of motion, and therefore
does not modify the equations of motion to this order.

What is important for us is that the results are unique at this order modulo field redefinitions. As long as the action agrees with results from string scattering computations (as checked most recently in~\cite{Chemissany:2007he}), it is determined by supersymmetry. This is known to be true also including terms of $O(\alpha'^2)$.

It should be possible to go beyond this order and  determine the exact string effective action including terms of $O(\alpha'^3)$ using the techniques of~\cite{Green:1998by, Basu:2008cf}. That would include various $R^4$ type couplings. Indeed there should be special couplings determined to all orders in the momentum expansion by space-time recursion relations.

For us, the tree-level result is sufficient. 
The conditions for unbroken supersymmetry follow from the
supersymmetry variations of the fermions of the ten-dimensional
effective action.  To lowest order in $\alpha'$, these are the
supersymmetry variations of ${\cal N}=1$ supergravity. The
space-time fermions consist of a gravitino, $\Psi_M$, which is a
Majorana--Weyl spinor. There is also a dilatino, $\lambda$, and a
gaugino $\chi$. Both are Majorana-Weyl spinors. The bosonic terms in
the supersymmetry variations of these fermions give the Killing
spinor equations that need to be satisfied: \be
\label{sugravariations}
\begin{split}
& \delta \Psi_M  =\Big(\partial_M +{1\over 4} {\Omega^{AB}_-}_M
\Gamma_{AB}\Big)\epsilon=0,\cr & \delta \lambda =-{1\over 2
\sqrt{2}} \Big(  /\!\!\!
\partial \hetdil  -{1\over 2}/\!\!\! \!{ \cal H} \Big) \epsilon=0,
\cr  &  \delta \chi  = -{1\over 2} \displaystyle{\not} {\cal F} \epsilon=0
,\cr
\end{split}
\ee where we have defined the following contractions of ${\cal H}$: \be
 /\!\!\! \!{\cal
H}_{M}= {1\over 2} {\cal H}_{MNP} \Gamma^{NP} \qquad {\rm and }
\qquad /\!\!\! \!{ \cal H}={1\over 3!} {\cal H}_{MNP} \Gamma^{MNP}.
\ee
With the convenient choice of fields considered in~\cite{Bergshoeff:1989de}, we note that all the modifications at $O(\alpha')$ to the supersymmetry variations of the space-time fermions are contained in the natural modification of ${\cal H}$ given in~\C{aaav}.

The Bianchi identity associated with the modified ${\cal H}$ of~\C{aaav}\ is \be\label{bianchiwconn} d{\cal H}
= { \alpha' \over 4}  \left\{
 \tr [R(\Omega_+) \wedge R(\Omega_+)] - \tr [F \wedge F] \right\}. \ee
If we wish to use the simple form of
the variations~\C{sugravariations}\ then there is a preferred
connection, $\Omega_+$, appearing in~\C{bianchiwconn}. This is going
to play an important role for us in understanding how duality
generates solutions of the Bianchi identity.

The supersymmetry conditions from the spinor variations~\C{sugravariations}\ can be
recast as conditions on the metric and fluxes of a torsional heterotic solution
summarized in~\cite{Strominger:1986uh}. The
torsional compactification requires a complex manifold with a
Hermitian $(1,1)$ class $J$ which plays a role analogous to the
usual K\"ahler form \be g_{a\bar{b}} = - i J_{a\bar{b}}. \ee The
torsion can be extracted directly from $J$ \be {\cal H} = {i}
\left( \partial -\bar\partial \right) J \ee with the K\"ahler case
corresponding to $dJ=0$. The dilaton satisfies the conformally balanced
condition
\be
d\left( e^{-2\hetdil} J \wedge J \right) =0.
\ee
The Bianchi identity~\C{bianchi}\ can be
expressed in terms of $J$ \be\label{bianchiH} d{\cal H} = 2 i
\partial\bar\partial J = { \alpha' \over 4}  \left\{ \tr [R(\omega)
\wedge R(\omega)] -\tr (F \wedge F) \right\} \ee which forces
$d{\cal H}$ to be a $(2,2)$ form. This is the only real constraint
from world-sheet supersymmetry on the choice of connection $\omega$
used in computing the Pontryagin class $\tr  (R \wedge
R)$~\cite{Sen:1986mg, Hull:1985dx}.

The difference between any two
choices of connection will yield an exact form given in terms of the
Chern-Simons invariant (CS) \be \tr [R(\omega_1) \wedge R(\omega_1)]
- \tr [R(\omega_2) \wedge R(\omega_2)] = d {\rm CS}(\omega_1,
\omega_2), \ee where
\begin{equation}
{\rm CS}(\omega_1,\omega_2) = 2 \alpha \wedge R(\omega_1) - \alpha
\wedge d \alpha -2 \alpha \wedge \omega_1 \wedge \alpha + { 2\over
3} \alpha \wedge \alpha \wedge \alpha,
\end{equation}
and $\alpha = \omega_1-\omega_2$. This term is of order $O(\alpha')$
in~\C{bianchiH}\ and so it might be possible to absorb the
difference in a redefinition of ${\cal H}$. Such a redefinition
would in turn correct $J$ and the resulting metric may or may not
satisfy the conditions for world-sheet conformal invariance.

From the
space-time perspective, we have already seen that the choice
$\omega = \Omega_+$ is special since it is compatible with space-time
supersymmetry giving a simple form for the spinor variations. Essentially supersymmetry
determines a preferred connection.

The way we will think about the torsional solutions -- certainly
those of DRS type and the generalizations we find here -- is in
terms of the original pre-duality type IIB supergravity metric and
fluxes. These quantities are unambiguous in the type IIB frame where
there is always a large volume limit. After duality, this data
determines the topology of the non-K\"ahler space. There is
additional subleading information suppressed by powers of $\alpha'$
which is needed for the complete solution but which is, however,
subject to the ambiguity of field redefinitions.

To complete our discussion of the background material, let us list
the dictionary relating type I and heterotic supergravity solutions.
Again there can be field redefinition ambiguities at higher orders
in $\alpha'$ beyond the supergravity data. The fluxes map in a
simple fashion \be {\cal H}_3 \, \leftrightarrow \, F_3'. \ee The
coupling constants and ten-dimensional metrics have the following
relations:
\begin{equation}
e^{\typeidil}  =e^{- \hetdil} \qquad {\rm and } \qquad  ds^2_I =
e^{\typeidil} ds^2_{het}
\end{equation}
where $\typeidil$ is the type I dilaton and $\hetdil$ is the
heterotic dilaton.

\subsection{The M-theory starting point}
\label{mstart}

The approach that we will use is to start with a consistent M-theory
compactification with $G_4$-flux. The data involved in this compactification
is a Calabi-Yau (CY) $4$-fold $\W$ and a choice of $(2,2)$ primitive $G_4$-flux. The
membrane tadpole
condition can be satisfied by a combination of flux and M2-branes~\cite{Becker:1996gj, Sethi:1996es}.

Generic $\W$ do not give rise to four-dimensional compactifications so we will insist
that $\W$ is elliptically-fibered with section. So there is a projection
\be \label{4fold} \pi_1 : \W \rightarrow {\mathbb B}_6 \ee
with torus fibers. For compatible choices of $G_4$-flux described in~\cite{Dasgupta:1999ss}, we can take the F-theory limit where the
volume of the
elliptic fiber goes to zero. We are left with type IIB on $ {\mathbb B}_6$ with a coupling constant $\tau_B$ determined by the complex
structure of the elliptic fiber. The $G_4$-flux will lift to a combination of type IIB fluxes
\be\label{defineG}
G_3 = F_3 + i e^{-\iibdil} H_3
\ee
together with gauge-bundles on the assorted $(p,q)$ $7$-branes present in this background. For a supersymmetric background, we impose the following condition~\cite{Dasgupta:1999ss}
\be\label{ISD}
\star G_3 = i G_3
\ee
along with primitivity with respect to the K\"ahler form $J$
\be
J \wedge G_3 =0.
\ee
These constraints relate the NS and RR fluxes
\be\label{nsrrrelations}
F_3 = \star \left( e^{-\iibdil} H_3 \right).
\ee

If we also consider spaces $\W$ that admit at least one $K3$
fibration then there should be a three-dimensional heterotic dual.
This is not the most general condition for a heterotic dual but it
will suffice for our discussion here. The most general condition has
yet to be formulated precisely but it is definitely more general
than a $K3$ fibration for three-dimensional compactifications; see,
for example,~\cite{Halmagyi:2007wi}. Nevertheless, we will assume a
second projection \be \pi_2 : \W \rightarrow  {\mathbb B}_4 \ee with
$K3$ fibers. If we want a four rather than three-dimensional
heterotic dual then we would like our $K3$ and elliptic fibrations
to be compatible so we can again take the F-theory limit.

Now the usual technique for constructing the heterotic dual is to replace the $K3$ fibers with
elliptic fibers. In doing so, we replace $\W$ with an elliptic Calabi-Yau $3$-fold $\M_H$ with base ${\mathbb B}_4$.
The data encoded
in the $K3$ fibration of $\W$ determines both how the elliptic fibers of $\M_H$ vary over ${\mathbb B}_4$ and the structure
of the heterotic gauge bundle.

This simple replacement makes sense when the F-theory
compactification only involves D3-branes and no flux. In this case,
the heterotic dual has been described in some
detail~\cite{Friedman:1997yq}. For compactifications with flux, the
dual heterotic geometry and bundle depend both on $G_4$ as well as
on $\W$. Even for a fixed $\W$, different choices of $G_4$ can give
rise to both torsional and non-torsional heterotic duals. Indeed the
most general heterotic dual is not a classical geometry but will be
a background that requires quantum patching conditions involving
T-duality. For those backgrounds, there are no sharp distinctions
between bundle and ${\cal H}$-flux.

The original DRS torsional solutions were obtained using a direct duality chain from F-theory on $K3\times K3$.
In this work, we will extend this construction to more general $\W$. Unfortunately, we need to
work with actual metrics rather than just complex geometry. Since little is known about explicit metrics for compact
Calabi-Yau spaces, we will use a semi-flat approximate metric to be described in section~\ref{semiflat}.

Before examining the implications of duality in a more detailed fashion, we can make some general comments about the structure of the
resulting heterotic solutions on general grounds. In the starting M-theory compactifications, the heterotic string is
realized by wrapping an M5-brane on the $K3$ fiber. The M5-brane supports a chiral $2$-form tensor $b_2$. Since the signature of the lattice $H_2(K3, \Z)$ is $(3,19)$, the Kaluza-Klein reduction of the M5-brane on the $K3$ fiber gives rise to $3$ compact scalars parametrizing $T^3$~\cite{Cherkis:1997bx}. In the F-theory limit where the elliptic fiber of the $K3$ is taken to zero size, $2$ of the $3$ scalars remain normalizable leaving a $T^2$.

What the choice of $G_4$-flux determines is the way in which this $T^2$ fibers vary over ${\mathbb B}_4$ and this determines the dual heterotic geometry. If the $G_4$-flux lifts strictly to gauge-bundle on $7$-branes in the F-theory limit  (i.e. the flux is localized on the discriminant locus of the elliptic fibration for $\W$) then the heterotic dual is a Calabi-Yau space. More generally the dual is non-K\"ahler with torsion. It is important to note that there are {\it many} dual geometries for a given $\W$ that depend on the particular choice of $G_4$-flux.  This is how a landscape-like degeneracy emerges for heterotic compactifications.

As pointed out in~\cite{Sethi:2007bw}, the wrapped M5-brane unifies
the various possible dual heterotic geometries and bundles in one
frame-work that depends on a choice of $K3$ fibration for $\W$ and a
choice of $G_4$-flux.  So the general form of the metric for the
dual heterotic geometry in all these cases will have the schematic
form \be\label{generalform} ds^2_{het} = ds^2_{base} + |dw + A|^2
\ee where the connection $A$ determines the structure of the torus
fibration and $w$ is a coordinate for the torus fiber.  That the
M5-brane captures both torsional and non-torsional geometries in the
same framework is almost an existence proof for heterotic geometries
of the form~\C{generalform}\ at least perturbatively.

\subsection{The type IIB orientifold locus}
\label{orientifoldlocus}

There is a class of elliptic CY $4$-folds $\W$ that admit a non-singular orientifold
locus where the elliptic fibration is locally constant. To make the problem of constructing
heterotic metrics more tractable,
we will consider this class.
At such a locus, we can
restrict our attention to type IIB on an elliptically-fibered
Calabi-Yau $\M$. As a complex space, we can express $\M$ in
Weierstrass form \begin{equation} \label{complex} {\tilde z} {\tilde y}^2 = {\tilde x}^3 + f {\tilde z}^2 {\tilde x} +
{\tilde z}^3 g \end{equation}
where $( {\tilde x}, {\tilde y}, {\tilde z})$ are homogeneous coordinates for $\CP^2$.

The choice of $(f,g)$ parametrizes the choice of
elliptic fibration. The symmetry
\begin{equation}
\label{quotient}  {\cal I}: \, {\tilde y}
\rightarrow - {\tilde y}
\end{equation}
acts as inversion on the elliptic
fiber. With such a symmetry, we can define a quotient of type IIB on $\M$ by the
symmetry $(-1)^{F_L} \,\cdot \Omega \, \cdot {\cal I}$.  Note that the quotient $\M/{\cal I}$ is ${\mathbb B}_6$ of~\C{4fold}.

We will need the string-frame metric of the type IIB compactification. This is a ten-dimensional warped metric of the form
\be
ds^2 = \Delta(y)^{-1} \, \eta_{\mu\nu} dx^\mu dx^\nu + \Delta(y) \, ds_{\M}^2(y)
\ee
where $\Delta(y)$ is the warp factor and we have chosen coordinates $y$ for the internal space $\M$. It is important to note that
there will be $\alpha'$ corrections to this metric. We only expect this form to be valid at large volume. We will return to this point
in section~\ref{tracking}.

Along with this warped metric will be $H_3, F_3$ fluxes along $\M$ and an $F_5$ flux with space-filling components. We will also take a general constant complex type IIB coupling $\tau_B$.

Both the $F_3$ and $H_3$ fluxes are odd under $(-1)^{F_L} \,\Omega$ so they must be
inverted by the $\Z_2$ action ${\cal I}$. This is rather critical because we
will later want to T-dualize both directions of the torus fiber.
We should note that there is a very large degeneracy of such type IIB
solutions from both the choice of $\M$ and the choice of fluxes. To proceed we need an explicit
form for the metric $\M$ suitable for T-duality.

\subsection{The metric for an elliptic CY space}
\label{semiflat}

Although we do not know the exact metric of an elliptic Calabi-Yau space
\be \pi: X \rightarrow {\mathbb B} \ee
with section, we can still express the metric in a semi-flat approximate form that makes the torus isometries of the elliptic fiber manifest. For this discussion, the dimension of the space $X$ can be general. The exact smooth metric has no such exact isometries and the breaking of the isometries occurs at the location of divisors in ${\mathbb B}$ where the elliptic fiber degenerates. These divisors can be viewed as supporting $(p,q)$ $7$-branes. For this reason, the semi-flat metric is related to the stringy cosmic string metric of~\cite{Greene:1989ya}. The approximate metric is a very good approximation to the actual metric with deviations that decay exponentially fast away from these divisors~\cite{gross-2000}.

We want to express the metric in a form that will allow us to later apply T-duality to both cycles of the elliptic fiber if we desire so the semi-flat form is ideal. In the absence of flux, this T-duality transformation should roughly send
\be
\tau \, \rightarrow -{1\over \tau}
\ee
where $\tau$ is the complex structure of the elliptic fiber. The metric is then invariant.

There is a
physical approach to the question of determining the semi-flat metric that goes as follows: consider type IIB on a $d$-dimensional complex base space with
coordinates $y$ and string-frame metric
\begin{equation}
ds_{\rm IIB}^2 = W(y) g_{ij} dy^i dy^j + W(y) \sum_{\mu=0}^{9-d} dx^\mu dx_\mu.
\end{equation}
We have allowed a warp factor $W(y)$ in front of the space-time metric which we will determine momentarily. The warp factor in front of $g_{ij}$ is for later convenience. This is a good solution of type IIB
string theory of F-theory type if the string coupling is determined by the complex structure of the elliptic fiber
\begin{equation}\tau (y) = \tau_1+i\tau_2= C + i e^{-\Phi}. \end{equation}
By definition, we assume this background satisfies the type IIB supergravity equations of motion with allowed sources given
by $(p,q)$ $7$-branes.

To get the metric of $X$, we compactify a spatial direction with coordinate $w_2$ and periodicity $w_2 \sim w_2 + 2\pi$. Let us T-dualize along this direction to get the IIA metric
\begin{equation}
ds_{\rm IIA}^2 =  W(y) g_{ij} dy^i dy^j + {1\over W(y)} (dw_2)^2 + W(y) \sum_{\mu=0}^{8-d} dx^\mu dx_\mu.
\end{equation}
There is no induced $B$-field but the dilaton becomes
\be
e^{2 \Phi_{IIA}} =  {1\over W(y)} e^{2 \Phi}.
\ee
The only other change is to the Ramond potential
\begin{equation}
C_1 = (C_0)_{w_2}.
\end{equation}
This is very clean. Next we lift the solution to M-theory which gives a metric for $X$. Let $w_1$ denote the eleventh direction with periodicity $2\pi$ then
\bea
ds^2_{\rm M} &=&  W^{4/3} e^{-{2\over 3} \Phi} \left(g_{ij} dy^i dy^j + {1\over W} (dw_2)^2 \right) + W^{-2/3} e^{{4\over 3} \Phi} \left( dw_1 -   (C_0)_{w_2} dw_2 \right)^2 \cr
&& + \, W^{4/3}  e^{-{2\over 3} \Phi} \sum_{\mu=0}^{8-d} dx^\mu dx_\mu .
\eea
Now we demand that this be a simple product M-theory solution so we impose the condition
\be W^{4/3}  e^{-{2\over 3} \Phi} = 1 \quad \Rightarrow \quad W = e^{\Phi/2}. \ee
The result is an approximate metric for an elliptic Calabi-Yau that exhibits isometries in
$(w_1,w_2)$ along which we can later T-dualize. There are quantum corrections to this metric which break the torus isometries
at the locations of the $7$-branes to which we will return later.

Let us express the semi-flat metric for $X$ in terms of the original geometric data,
\be\label{metricM}
ds^2_{X} =   \, g_{ij} dy^i dy^j  +
{1 \over \tau_2} |dw_1 - \tau \,  dw_2|^2.
\ee
Note that we have not exhibited a complex structure for $X$ although we could certainly take complex coordinates for the
base space. In terms of those complex coordinates, $\tau$ is holomorphic away from the discriminant locus where the elliptic
fiber degenerates.

\section{Torsional Solutions via Duality}

Now we want to go from our starting point of type IIB on $\M$ to a
heterotic geometry. How shall we proceed? There are five cases where
we can do this rigorously of which only the original construction
of~\cite{Dasgupta:1999ss}\ using $\M = K3 \times T^2$ has been
studied to date. The remaining four cases correspond to special
three-folds $\M$ which {\it themselves} admit a non-singular
orientifold limit. Those special spaces, denoted $\M_n$, are
elliptic fibrations over Hirzebruch surfaces $F_n$ for the choices
$n=0,1,2$ and $4$~\cite{Sen:1997gv, Sen:1997bp}.\footnote{When used
to construct six-dimensional F-theory compactifications, the cases
$n=0$ and $n=2$ are equivalent since we are allowed to consider
generic moduli~\cite{Morrison:1996pp}. It is unlikely that this is
true in the current context where we are constructing
four-dimensional compactifications that depend critically on the
choice of flux. In particular, the flux lifts moduli so we are no
longer free to deform
  to
generic metrics.}

So our approach will be to dualize each of the cases that admit orientifold limits. By doing so, we will find
generalizations of the original DRS geometries which we can check satisfy the constraints for a torsional heterotic compactification. However, heterotic/F-theory duality discussed in section~\ref{mstart}\ strongly suggests that there should exist torsional heterotic compactifications associated to any elliptic $3$-fold $\M$.
So we will actually discuss more general elliptic three-folds $\M$ keeping in mind that the duality is rigorous for the special spaces mentioned above. Despite the absence of a rigorous duality in the general case, we should stress that the duality procedure gives good solutions in the absence of flux and we expect it to give good solutions in the presence of flux.

\subsection{The Method of Construction}
\label{method}

Now we want to take the metric~\C{metricM} as the starting point for
a type IIB compactification on $\M$. The type IIB coupling, $\tau_B
= C_0 + i e^{-\iibdil}$, is independent of the coordinates for
$\M$. 
Initially there is no $B$-field. First note that if we were to
T-dualize the metric~\C{metricM}\ along the $(w_1,w_2)$ directions,
we would not generate any $H$-field. This is natural. We expect a
purely geometric NS string background to map to a purely geometric
NS background.

To find new torsional solutions, we must also include $H_3, F_3$ and $F_5$ in our initial background. As a first step, we want to determine the torsional metric so we need only consider the NS sector $H_3$-flux. Our starting ten-dimensional IIB metric is given by the semi-flat approximation to the Calabi-Yau metric for $\M$,
\be \label{preduality}
e^{-{\iibdil\over2}} ds^2_{IIB} = \Delta^{-1} \eta_{\mu\nu} dx^\mu dx^\nu + \Delta \left(   g_{ij}(y) dy^i dy^j  + {1 \over \tau_2} |dw_1 - \tau \,  dw_2|^2 \right),
\ee
with an additional warp factor $\Delta$. The warp factor is related to space filling $C_4$ potential~\cite{Dasgupta:1999ss}
\be
(C_{4})_{ \mu\nu\rho\lambda} = {1\over \Delta^2 }\, \epsilon_{\mu\nu\rho\lambda}.
\ee
The base obtained by projecting
\be
\pi: \M \rightarrow  {\mathbb B}_4
\ee
has metric $\Delta g_{ij}$ and the elliptic fiber degenerates over divisors of ${\mathbb B}_4$. Over these loci, we expect the smooth metric for $\M$ to differ from the semi-flat approximation.

We need to express the $G_3$-flux in a form suitable for duality. Note that we can decompose differential forms on $\M$ under the action of the $\Z_2$ involution ${\cal I}$ given in~\C{quotient}. Our interest resides in $3$-forms
\be
\Omega^3 (\M) = \Omega^3_+ \oplus \Omega^3_-
\ee
to which we can assign a definite charge under ${\cal I}$. Now let us consider either the Ramond or NS type IIB $3$-form field strength
which is a form
\be
f_3 \in H^3(\M, \Z)_-
\ee
invariant under the orientifold action as described in section~\ref{orientifoldlocus}.

Because of the $\Z_2$ action on forms, we can nicely decompose $f_3$ along the fiber and base as follows: to the integral form $f_3$, we associate integral $2$-forms, $(f_2)_i$, living on the base $B_4$ of the elliptic fibration with values in the cohomology of the fibers of degree $1$. Each form $f_2$ is well-defined on the base up to the action of the $SL(2,\Z)$ monodromy group that acts on the $1$-forms of the fiber when we consider  loops enclosing $7$-branes; equivalently, loops enclosing the divisors where the elliptic fibers degenerate.

If we choose $(\omega^1, \omega^2)$ as a basis for the integral
harmonic $1$-forms of the fiber then \be f_3 = (f_2)_i \wedge
\omega^i. \ee Under the $SL(2,\Z)$ action sending \be \tau \,
\rightarrow \, {a\tau + b \over c \tau +d}, \ee the combination \be
{\tilde f}_2 = (f_2)_2 + \tau (f_2)_1 \ee transforms as a modular
form of weight $(-1,0)$: \be {\tilde f}_2 \, \rightarrow \, (c\tau +
d)^{-1} {\tilde f}_2. \ee Let us consider a patch in which the
action of the monodromy group is trivial. In such a patch, we can
trivialize each $(f_2)_i$ to obtain two $1$-form connections again
with values in the cohomology of the fibers of degree $1$ \be
(f_2)_i = d (A_1)_i. \ee So, for example, we can take $f_3= H_3$. In
this case, the two connection $1$-forms correspond to trivializing
$B_2$ in terms of two local potentials which we denote \be \bwone =
B_{y_iw_1} dy^i, \quad  \bwtwo = B_{y_iw_2} dy^i \ee using the
fiber/base coordinates for the metric~\C{preduality}. In terms of
these potentials, we define \be\label{connection1} A_H(\tau) =
\bwtwo + \tau \bwone \ee which is a $1$-form connection constructed
from the NS field strength using this procedure.


We now have enough information to apply T-duality to both directions $(w_1, w_2)$ of the elliptic fiber. After applying T-duality in these directions, we arrive at a new space $\M'$ with the metric~\C{preduality}\
becoming
\bea \label{wwwx}
e^{-{\iibdil\over2}} ds^2_{\rm tor} &=&  \Delta^{-1} \eta_{\mu\nu} dx^\mu dx^\nu +  \bigg( \Delta\,  g_{ij} dy^i dy^j  \, +\cr &&  {e^{-{\iibdil}} \over \Delta \tau_2} |dw_2 + \tau \,  dw_1 + A_H |^2 \bigg).
\eea
There is no $B$-field generated in the final solution as we desire.
After this duality, we have arrived at the NS geometry of type I compactified
on non-K\"ahler torsional metrics that generalize the earlier known solutions.

There are a few points to mention. Whether the metric exists depends on whether there is an obstruction to
finding a suitable warp factor $\Delta$. We will argue in section~\ref{tadpole}\ that this obstruction is the tadpole
cancelation condition in type IIB which becomes the $5$-brane tadpole cancelation in type I or heterotic string theory. This can be satisfied in type IIB for suitable fluxes which gives the warp factor $\Delta$.

The second key issue is whether the singularities of the elliptic fiber of this semi-flat metric can be smoothed to give a good torsional solution. We expect this to be the case since the metric can be smoothed in the type IIB frame. The semi-flat approximation is an extremely good approximation to the actual smooth metric with only exponentially small corrections near the degeneration divisors.

Including these small corrections will break both isometries of the elliptic fibration in a way familiar from other examples of T-dualizing approximate isometries; for example, the duality between NS5-branes on a circle and Taub-NUT spaces. We will discuss aspects of these quantum corrections and the desingularization procedure in section~\ref{desing}. In complex dimension three of relevance to~\C{wwwx}, little can be said explicitly about the smoothing and a rigorous existence theorem is needed except for the orientifold cases described at the beginning of this section.

It is going to be convenient for us to express the torsional internal metric~\C{wwwx}\ in terms of a flat frame,
\be
ds^2_{\rm tor} = \sum_{a=1}^4 e^{{\iibdil\over2}} E^a E^a+ e^{-{\iibdil\over2}} E^w  E^{\bar w}.
\ee
We have defined the orthogonal frame as follows:
\be\label{appb}
E^a  =  \sqrt{\Delta} e^a, \quad
E^w  =  {1\over \sqrt{\Delta\tau_2}} (dw_2 + \tau dw_1 + A_H), \quad
E^{\bar w} =  (E^w)^\star. \,
\ee
The $e^a$ are vielbeine  for the unwarped base metric $g_{ij}$.

Finally, we also need to specify the type I dilaton which follows from the standard rules of T-duality
\begin{equation}
e^{\typeidil}={e^{{\iibdil \over 2}}\over \Delta(y)}.
\end{equation}
To avoid cluttering our subsequent formulae, we will now set the constant $e^{{\iibdil}}=1$.

\subsection{The RR fluxes} \label{rrfluxes}

To complete the description of the type I background, we need to specify the RR fluxes. The initial RR type IIB potentials take the form
\be\label{startingpotentials}
C_0, \quad C_2, \quad C_4,
\ee
where $\tau_B = C_0 + i e^{-\iibdil}.$ The field strength
\be
F_3 = dC_2 + H_3 \wedge C_0
\ee
is again odd under ${\cal I}$ so we obtain a second potential $A_F(\tau)$ on ${\mathbb B}_4$ with good transformation properties under the $SL(2,\Z)$ acting on the fibers. Now the potential $A_F$ is not really independent of $A_H$. The two potentials are connected via the imaginary self-duality condition~\C{ISD}. 

Type I string theory has one RR potential $C_2'$. We want to express
the potential that results from duality in a form that makes its
connection with the geometry~\C{wwwx}\ clear. The easiest terms to
consider in $C_2'$ are those proportional to $C_0$ which can be
expressed in terms of the $1$-forms~\C{appb}\ as follows, \be
\label{firstflux} C_2' = -{i \over 2} C_0 \Delta E^w E^{\bar w} +
\ldots. \ee There are two additional contributions from $(C_2, C_4)$
which need to be treated together. In a form convenient for
T-duality, the type IIB $5$-form field strength is given in terms of
a potential $C_4$ by \be \label{deff5} F_5 = d C_4 + H_3 \wedge C_2.
\ee The implied Bianchi identity includes the $H_3$-induced source
term for D3-brane charge, \be\label{abi} dF_5 = F_3 \wedge H_3. \ee
With the definition~\C{deff5}, $F_5$ is self-dual and so the Bianchi
identity is also the equation of motion for $C_4$. We will need to
add brane/orientifold sources to~\C{abi}. These sources come about
as follows: in M-theory, there is a higher derivative coupling \be -
\int C_3 \wedge X_8(R) \ee where $X_8(R)$ is constructed from
curvature tensors and given by the following combination of
Pontryagin classes \be X_8 = {1\over 48} \left( p_2 - {1\over 4}
p_1^2 \right). \ee This gives rise to an M2-brane tadpole given by
$\int_\W X_8$ when evaluated on a $4$-fold internal space ${\cal
W}$. If the $4$-fold is Calabi-Yau then the tadpole is $\chi (\W)
/24$.

On taking the F-theory limit, this coupling is reproduced by
gravitational couplings on $(p,q)$ $7$-branes. The branes wrap
divisors of the IIB compactification space $B$.  The gravitational
couplings for a brane wrapping a divisor $D$ is of schematic type
\be - \int C_4 \wedge X_4 \ee where $X_4 (D)$ is again constructed
from curvature tensors. From the bulk perspective, we can view this
as a $6$-form class added to the Bianchi identity~\C{abi} \be X_6
(D) = X_4 \wedge \delta^{2}_D \ee where the delta-function restricts
to the divisor $D$. The tadpole can be expressed purely in terms of
classes on $B$~\cite{Sethi:1996es} \be {\chi\over 24} = \int_B\left(
15 c_1^3 + {1\over 2} c_1 c_2 \right). \ee In the orientifold limit
where the IIB compactification space $B$ is itself a quotient of a
Calabi-Yau $\M$, there are only D7-branes and orientifold planes.
Each D7-brane wrapping a divisor $D$ supports a coupling, \be - \int
C_4 \wedge {1\over 24} c_2, \ee while the orientifold planes support
couplings~\cite{Dasgupta:1997cd}, \be - \int C_4 \wedge {1\over 6}
c_2. \ee For a recent discussion of the orientifold limit tadpoles,
see~\cite{Plauschinn:2008yd}. For our purposes, we can lump all
these contributions together into a class denoted, ${\tilde X}_6$,
which includes all the brane and orientifold modifications to the
Bianchi identity \be\label{modbianchi} dF_5 = F_3 \wedge H_3 +
{\tilde X}_6. \ee We will analyze how each contribution
to~\C{modbianchi}\ maps under duality in section~\ref{tadpole}.

After dualizing $( C_2, C_4)$ in addition to~\C{firstflux}, we can finally express the resulting type I $F_3'$ in a form
which will be convenient for later discussions
\begin{equation}\label{typeIflux}
F_3'= \sqrt{{\Delta} \over {\tau_2}} \, {\rm
Im} \left[ (F_{w_2} + \tau F_{w_1} ) E^{\bar w})\right] +\star_b d \Delta^2 .
\end{equation}
The Hodge star action, $\star_b$, is with respect to the unwarped
base metric $g_{ij}$. As we will later see when discussing the
spinor equations for supersymmetry, the first contribution
to~\C{typeIflux}\ combines with some components of the spin
connection to give rise to an self-dual $3$-form. This self-duality
is a direct consequence of the imaginary self-duality of the type
IIB $G_3$-flux.

\subsection{Tracking the volume moduli}
\label{tracking}

Now the original IIB flux compactification always has a physical modulus that corresponds to rescaling the internal six-dimensional IIB metric,
\be\label{rescaling}
{ds^2_{IIB}}  \, \rightarrow \, (L)^2 ds^2_{IIB}.
\ee
This is the only real tunable parameter in the IIB theory since the string coupling is typically frozen by the flux at some value. When $L$ is large, supergravity becomes a more reliable approximation as $\alpha'$ corrections are suppressed. There are actually at least two distinguished moduli for a metric of the form~\C{preduality}\ corresponding to independent scalings of the fiber and base.

We can keep track of the second modulus by also permitting separate scalings of the base metric,
\be
g_{ij}  dy^i dy^j \, \rightarrow \, ({\widetilde L})^2 g_{ij} dy^i dy^j .
\ee
For the moment, we need only keep track of the basic symmetry~\C{rescaling}. Physical moduli cannot disappear under a duality transformation. In type I, the rescalings above act on the metric in the following way
\bea
    \Delta\,  g_{ij} dy^i dy^j   \,  &\rightarrow & \,
  \Delta' \,  (L)^2 g_{ij} dy^i dy^j  \\
     {1 \over \Delta \tau_2} |dw_2 + \tau \,  dw_1 + A_H |^2 \, & \rightarrow & \,
   {1 \over  (L)^2 \Delta' \tau_2} |dw_2 + \tau \,  dw_1 + A_H |^2
\eea
together with an action on the type I dilaton
\be
\typeidil \, \rightarrow \, {\typeidil} - 2 \log (L).
\ee
The scaling of the warp factor is determined from~\C{modbianchi}\ together with the self-duality condition $ F_5 = \ast F_5$. Factoring out the powers of $L$, the warp factor obeys an equation of the form
\be\label{warpfactoreq}
d \ast d \left({1\over \Delta^{'2}} \epsilon_{\mu\nu\rho\lambda} dx^\mu \cdots dx^\lambda \right) = O({1\over L^4})
\ee
where the Hodge star is with respect to the $L=1$ metric; so the warp factor becomes more constant as $L \rightarrow \infty$.

The choice of $L$ parametrizes a family of solutions in type I (and consequently the heterotic string). However, the simplifying limit $L\rightarrow \infty$ corresponds in type I to the area of the torus fiber becoming small and the string coupling becoming small.

\subsection{The type IIB and type I tadpole conditions}
\label{tadpole}

\subsubsection{The supergravity contribution}

Now we would like to understand how a solution of the type IIB
D3-brane tadpole condition maps to a solution of the type I D5-brane
tadpole, and consequently the heterotic Bianchi identity. There are
two distinct contributions to~\C{modbianchi}\ that we will treat in
turn. We begin with the supergravity source term
\be\label{sugracontribution} dF_5 = F_3 \wedge H_3. \ee This must
map consistently under T-duality simply in type IIB string theory
whether or not we choose to orientifold or insert branes. In fact,
if we started with a non-compact space $\M$ in type IIB, we do not
even have to add extra ingredients to cancel the induced D3-brane
charge \be Q = \int F_3 \wedge H_3. \ee Let us apply T-duality in
the fiber directions to obtain the type I $3$-form \be
F_3'=(F_5)_{w_1 w_2} + (F_3)_{w_1} ( dw_2 + B_{w_2}) - (F_3)_{w_2} (
dw_1 + B_{w_1}). \ee The notation $(F_5)_{w_1 w_2}$ refers to the
$3$-form obtained by taking the component of $F_5$ proportional to
the volume form of the fiber, and removing that volume form to
obtain a $3$-form. There is a similar definition for $(F_3)_{w_i}$
given in Appendix~\ref{notation}.

At the level of supergravity the type I Bianchi identity then
becomes \be d F_3'=0. \ee Note that $F_3'$ is globally defined and
as a result $d F_3'$ is exact. However, even though $dF_3'$ is exact
there is still associated D5-brane charge. The charge cannot simply
vanish. The reason there is charge is that T-duality changes the
topology of the space and in the type I geometry the base of the
fibration now has a boundary. As a result, it is not a cycle. When
integrated over the base $dF_3'$ becomes $Q$, or in other words the
D3-brane charge which is induced by flux on the type IIB side is
generated by geometry in type I.

This is natural since, by construction, we chose T-dualities to precisely map $H_3$ into metric both in our examples and the earlier ones of~\cite{Dasgupta:1999ss}.  Let us illustrate this idea in a simple example.
Consider a $3$-torus represented by the product of three circles
\be
ds^2 = dx_1^2+dx_2^2+dx_3^3,\qquad x_i \sim x_i +1,
\ee
in the presence of $N$ units of NS $3$-form flux $H_{NS}$ which locally we can
trivialize by a $2$-form $B_{NS}$,
\be
H_{NS}=d B_{NS} = N dx_1 \wedge dx_2 \wedge dx_3  \quad {\rm with } \quad B_{NS}= N x_1 dx_2 \wedge dx_3.
\ee

Now apply T-duality in the $x_3$-direction which corresponds to one of our fiber directions (recall that the $H_3$-flux is odd on the fiber as described in section~\ref{orientifoldlocus}). The metric becomes a circle fibred over
a $2$-torus
\be
ds^2 = dx_1 ^ 2 + dx_2^2 + (dx_3- N x_1 dx_2)^2.
\ee
In order for the metric to be globally defined the boundary conditions on the coordinates have to
be changed
\be
x_i \sim x_i + 1 , \quad i=2,3 \qquad {\rm and } \qquad x_1\sim x_1 + 1 , \quad x_3 \sim x_3 + N x_2.
\ee
This change in boundary conditions implies that the topology of the space has changed.

Indeed, while the $3$-torus has betti numbers $b_0=b_3=1$ and
$b_1=b_2=3$, the T-dual space has betti numbers $b_0=b_3=1$ but
$b_1=b_2=2$. The reason for the change is that in the T-dual space,
the form
$$\omega= dx_3 - N x_1 dx_2$$
is globally defined. As a result its exterior derivative is exact
and $dx_1 \wedge dx_2$ becomes trivial in cohomology. Moreover,
$b_1$ is also changed since $\omega$ is no longer closed. By
Poincar\'e duality this implies that the base has a boundary. We may
integrate an exact form over the base of the fibration and obtain a
non-vanishing result. In particular \be \int _{base} d \omega = N.
\ee This is precisely what happens in our examples and the earlier
examples of~\cite{Dasgupta:1999ss}; namely, that the D3-brane charge
$Q$ induced by the fluxes in type IIB appears as five-brane charge
in the T-dual geometry where \be \int_{base} dF_3'=Q. \ee

\subsubsection{The gravitational contribution}

The supergravity contribution to the charge maps nicely as explained above. We would now like to turn to the tadpole contribution
that comes about from the gravitational couplings on the branes and orientifold planes denoted ${\tilde X}_6$ in~\C{modbianchi}. This is a more mysterious because it involves higher momentum couplings so there is room for possible quantum corrections.

Let us overview how this should work before jumping into a computation. In the heterotic frame we expect ${\tilde X}_6$ to map to the contribution,
\be
\alpha'  \tr [R(\Omega_+) \wedge R(\Omega_+)],
\ee
in the heterotic Bianchi identity~\C{bianchiwconn}\ up to an overall numerical factor. The curvature is evaluated with respect to the $\Omega_+$ connection which depends on ${\cal H}_3$. If we apply S-duality to convert heterotic to type I, we replace ${\cal H}_3$ by $F_3'$.  So in type I, we expect ${\tilde X}_6$ to map to a $4$-form proportional to $ \tr [R(\Omega_+') \wedge R(\Omega_+')]$ where
\be\label{typeIconn}
{\Omega_\pm}'= \Omega\pm \half {{
F}_3'}\left( 1 +O(\alpha') \right).
\ee
The omitted terms can be very complicated since they need not be linear in the fluxes.

The gravitational couplings on branes and orientifolds, ${\tilde X}_6$, are believed to be T-duality invariant using general arguments from K-theory in the presence of $H_3$ which is a pure torsion class~\cite{Witten:1998cd}. A precise general statement is that the $\alpha'$-corrected equations of motion including these couplings should be invariant under T-duality.

This does not determine which connection is to be used in computing these couplings. The choice of connection depends on which interactions are shuffled into bulk equations of motion and which into these gravitational couplings. This is ambiguous. Perturbative string computations suggest that the $H_3$-connection is preferred over the metric connection; steps toward showing this appear in~\cite{Scrucca:2000ae}.

What we require is actually something additional: namely, dependence on the RR field strengths as well as NS field strengths. The appearance of both RR and NS fluxes in anomaly cancellation is already visible in M-theory~\cite{Lukic:2007aj}.  This kind of dependence is also needed, in part, to ensure the equivalence of the type I and heterotic string under S-duality even in ten dimensions.

Again we stress that the appearance of the $H_3$-connection in heterotic or the $F_3$-connection in type I is based on a nice form for the equations of motion and the supersymmetry transformations. In this nice choice of fields, the first corrections to the spinor variations~\C{sugravariations}\ are at $O(\alpha'^2)$ with no corrections at $O(\alpha')$. So it makes little sense to try to solve the heterotic or type I Bianchi identity beyond this order in the $\alpha'$ expansion since the metric and fluxes will be corrected. However, we do expect to find a solution from duality to this order.

Now we have framed this discussion from the heterotic/type I perspective. We should ask a similar question about the order of quantum corrections in type IIB. Without orientifolding or branes, the leading corrections would be more suppressed generated by terms in the ten-dimensional effective action like $R^4$ which is down by $O(\alpha'^3)$ from the supergravity terms. However, the orientifolded theory with branes has quantum corrections to the metric and fluxes at precisely the same order as type I/heterotic.

What is true on the type IIB side is that the warp factor equation~\C{warpfactoreq}\ is of Laplace type together with higher derivative corrections. The only obstruction to the existence of a perturbative solution comes from the zero mode for the sources, which is the usual tadpole condition. Once that condition is satisfied a solution exists.

We therefore want to show that the candidate warp factor $\Delta$ and associated type I flux $F_3'$ of~\C{typeIflux}\ obtained from the type IIB side by T-duality satisfy the type I Bianchi identity to leading orders in the type IIB large volume expansion when we evaluate curvatures using~\C{typeIconn}.

So our task is to evaluate $p_1$ for the torsional compactification
metric~\C{wwwx}. We will be interested in the leading order
result in the $L$ expansion described in section~\ref{tracking}. This is an expansion around the limit of large base and small fiber
for the torsional metric. Constructing an obstruction theory to a perturbative solution is natural in this expansion which is T-dual to the
large volume expansion of type IIB.

The first step
is to evaluate the spin connection for the vielbeine given
in~\C{appb}. We will need the spin connection, $\omega$, for the
underlying Calabi-Yau metric $g_{ij}$. Let us define $\Delta_b$
via \be d \log \Delta = \Delta_b E^b. \ee The spin
connection for the torsional metric is given by
\bea {\Omega^a}_b &= & {\omega^a}_b-{1\over 2} (\Delta^{a} E_{bj}-\Delta_b
E^a_j ) dy^j- {1\over 2\sqrt{\tau_2 \Delta} } E^{ai}E^j_b \, {\rm Re}
\left[ \left(H_{w_2} + \bar \tau H_{w_1} \right)_{ij}E^w \right]\cr
{\Omega^w}_a  &= & {i\over 2 \tau_2} E^i_a
\partial_i \tau E^{\bar w} +{1\over 2\sqrt{\tau_2 \Delta} } (H_{w_2}+ \tau
H_{w_1}) _{ij} E^i_a dy^j-{1\over 2} \Delta_a E^w\cr {\Omega^w}_w &=
& {i \over 2 \tau_2} d \tau_1.
\eea
Using the scalings described in section~\ref{tracking}\ shows that the
$\Omega_+$ connection coefficients satisfy
\bea {\Omega_+^a}_b &= & {\omega^a}_b+O(L^{-2}), \cr {\Omega^w_+}_a &= & O(L^{-1}),
\cr {\Omega^w_+}_w &= & {i \over 2 \tau_2} d \tau_1.
\eea
Using this expansion to evaluate the curvature correction gives,
\begin{equation}\label{bianchiexpand}
{\rm tr} \left[ R(\Omega_+) \wedge R(\Omega_+) \right] = {\rm
tr}\left[ r \wedge r\right]+O(L^{-1}),
\end{equation}
where $r$ denotes the curvature $2$-form of the base. Note that even
though there is one spin connection depending on $\tau$ which is
$O(1)$ for large $L$, namely $\Omega^w_{+w}$, it does not contribute
to the right hand side of~\C{bianchiexpand}\ to leading order in $L$.

In the large $L$ limit,  the Bianchi identity then reduces to
\begin{equation}
d \star_b d \Delta^2+ \left( F_{w1} \wedge H_{w2} - F_{w2} \wedge H_{w1} \right)= {\alpha'\over 4 } \tr
\left[r\wedge r\right] + O(L^{-1}).
\end{equation}
At $O(L^2)$, this equation states the the warp factor has a constant
piece. At $O(1)$, we obtain a differential equation for the warp
factor of Laplace type with a source.
This equation will always have a solution as long as the source has no zero mode. This is
the statement that the NS5-brane charge vanish.

This is what we wanted to see. We can recognize
this as the warp factor equation on the type IIB side, where the
$\tr [ r\wedge r]$ piece arises from an anomalous coupling on the D7
brane world-volume wrapping the base of the elliptic fibration as described in
section~\ref{rrfluxes}. The
integrability condition is the tadpole cancelation condition. It is
possible to analyze the Bianchi identity beyond the first two orders
of the $L$ expansion considered here. On the type IIB side, this corresponds to
corrections to the warp factor equation. A detailed analysis of the
Bianchi identity beyond the first two leading orders for the case where the base is $K3$ will
appear in~\cite{toappearbecker}.


This is not a complete analysis of the Bianchi identity for the varying $\tau$ case. In that case, there can be extra
contributions from $4$-cycles associated to degenerations of the elliptic fiber which require additional
analysis. At this point, however, we can be more confident that a warp factor which works in type IIB
will define a good heterotic or type I background in the perturbative expansion  that we have described.






\subsection{Checking the supersymmetry conditions}
\label{checkingSUSY}

\subsubsection{Semi-flat metric with one holomorphic parameter}

We now turn to the supersymmetry properties of our backgrounds. Let us
begin by showing that one ten-dimensional
Majorana-Weyl spinor compactified on a space with semi-flat metric~\C{metricM}\ gives N=1 supersymmetry in four
dimensions. Our analysis is local since the supersymmetry conditions must be satisfied point-wise. This will allow
us to work in patches avoiding singularities and monodromies of the complex structure of the elliptic fiber.

Take a ten-dimensional Majorana-Weyl spinor,
$\epsilon$, which we choose to have positive chirality
\begin{equation}
\Gamma_0 \cdots \Gamma_9 \epsilon=\epsilon.
\end{equation}
On compactifiation to four dimensions, the
Majorana-Weyl spinor $\epsilon$ decomposes into a
four-dimensional complex Weyl spinor, $\zeta$, and a six-dimensional
complex Weyl spinor $\xi$:
\begin{equation}
\epsilon
= \zeta \otimes \xi + \zeta^\star \otimes \xi^\star.
\end{equation}
Decomposing the ten-dimensional Dirac matrices into four- and
six-dimensional pieces, \be \Gamma_{M} = \gamma_5 \otimes
\gamma_M\qquad {\rm and } \qquad \Gamma_{\mu} = \gamma_\mu \otimes
1, \ee gives rise to spinor equations in six dimensions
\begin{equation}\label{kkki}
\delta \Psi_M = \nabla_M\xi=\partial_M \xi + {1\over 4}
{\Omega^{AB}}_M\gamma_{AB} \xi=0.
\end{equation}
The non-vanishing connections are
\begin{equation}
\begin{split}
& {\Omega^a}_b={\omega^a}_b, \cr &  {\Omega^w}_a = -{1\over 2 i
\tau_2} {e^i}_a
\partial_i \tau E^{\bar w},\cr & {\Omega^w}_w={i\over 2 \tau_2} d
\tau_1,
\end{split}
\end{equation}
together with their complex conjugates. The spin connection is an
$SO(6)$ gauge field
and therefore each {\bf 4} of $SO(6)$ can give rise to one singlet
under the holonomy group, resulting in at most minimal supersymmetry
in four dimensions.

To preserve supersymmetry, the spinor equations $ \delta \Psi_M =0$
must be satisfied. This requires the space to be K\"ahler since the
$2$-form
\begin{equation}
J_{MN}=-i \xi^\dagger \gamma_{MN}\xi,
\end{equation}
is covariantly constant and satisfies $J^2=- {\bf 1}$. Let us introduce
complex coordinates for the six-dimensional space and
choose $\xi$ to satisfy
\begin{equation}
\gamma_{\bar M}\xi=\gamma^{M}\xi=0.
\end{equation}
To solve~\C{kkki}, $\tau(y)$ has to be an holomorphic function
of the base coordinates. This follows from,
\begin{equation}
\delta\psi_{w_1} = \partial_{w_1} \xi + { i \over 8} \tau_2^{-3/2}
\bar \partial_{\bar i} \tau {\gamma_{w}}^{\bar i} \xi=0,
\end{equation}
which requires $\xi$ to both be independent of the $w_1$ coordinate and satisfy
\begin{equation}
\bar \partial \tau(y)=0.
\end{equation}
Now this analysis is only true away from singularities of the elliptic fiber. As we mentioned earlier, we are restricting our analysis to the locus omitting those singularities which means that we cut out a divisor from the six-dimensional space.

The spinor equation in the $w_2$ direction is solved by requiring
additionally that $\xi$ be independent of $w_2$. Vanishing of the gravitino variation
along the base, on the other hand, requires
\begin{equation}\label{kkii}
\delta \psi_i = \partial_i \epsilon + {1\over 4} {\omega^{ab}}_i
\gamma_{ab}\epsilon-{i \over 4} {\partial_i \tau_1\over \tau_2}
\epsilon=0.
\end{equation}
The integrability condition to find a solution of~\C{kkii}\ is
that the Ricci-form of the base is related to $\tau$ according to
\begin{equation}\label{kkiii}
{\cal R}=i\partial \bar \partial \log \tau_2 ,
\end{equation}
or since the base is K\"ahler
\begin{equation}
\partial \bar \partial\left(  \det \log g_{i \bar j } - \log \tau_2\right) =0.
\end{equation}

Thus compactification of a ten-dimensional Majorana-Weyl spinor on a
space with metric~\C{metricM}\ gives rise to one four-dimensional
supersymmetry if $\tau(y)$ is a holomorphic function of the base
coordinates and if the integrability condition~\C{kkiii}\ is
satisfied.

\subsubsection{Torsional heterotic background with one holomorphic parameter}
\label{toroneparameter}

The gravitino variation appearing in~\C{sugravariations}\ has a
nice interpretation as implying the existence of a covariantly
constant spinor when ${\cal H}$ is included. So let us examine how
the spin connection has changed in going from our initial
conformally Calabi-Yau metric to the torsional solution.

The torsional geometry in the heterotic frame is characterized by the metric
\begin{equation}
ds^2_{\rm tor} = \Delta^2\,  g_{ij} dy^i dy^j  + {1 \over \tau_2}
|dw_2 + \tau \,  dw_1 + A_H |^2.
\end{equation}
The base is K\"ahler with metric $g_{ij}$ and the flux is
\begin{equation}\label{repeatH}
{\cal H}= {1 \over \sqrt{\tau_2}} \, {\rm Im} \left[ (F_{w_2} + \tau
F_{w_1} ) E^{\bar w}\right] +\star_b d \Delta^2 .
\end{equation}

In order to analyze the supersymmetry constraints, we begin by
computing the $\Omega_-$ connection which appears in the gravitino
supersymmetry variation. We can take as a basis of $1$-forms
\begin{equation}
E^a = \Delta {e^a}, \quad E^w  = {1\over \sqrt{\tau_2}} \left( dw_2
+ \tau dw_1 +A_H\right), \quad (E^w)^\star,
\end{equation}
where $e^a$ is the orthonormal basis of the base. In terms of this
basis, the torsional metric is flat
\begin{equation}
ds^2 = \sum_{a=1}^4 E^a E^a+ E^w  E^{\bar w},
\end{equation}
up to an overall constant factor that we will set to one for
simplicity. The spin connection coefficients are
\bea {\Omega^a}_b &= & {\omega^a}_b+(E^i_b E^a_j
-E^{ai} E_{bj}) \left({\partial_i \log \Delta} \right)\,  dy^j -
{1\over 2}{\tau_2^{-1/2}} E^{ai}E^j_b \times \cr && {\rm Re} \left[ \left(H_{w_2} +
\bar \tau H_{w_1} \right)_{ij}E^w \right]\cr {\Omega^w}_a  &= &
{i\over 2 \tau_2} E^i_a
\partial_i \tau E^{\bar w} +{1\over 2}{\tau_2^{-1/2}} (H_{w_2}+ \tau
H_{w_1}) _{ij} E^i_a dy^j\cr {\Omega^w}_w &= & {i \over 2 \tau_2} d
\tau_1. \eea
To check the supersymmetry constraints, we need the $\Omega_-$
connection. Using~\C{repeatH}\ for ${\cal H}$ gives
\bea
{\Omega^a}_{-b}  &= &~ {\omega^a}_b+(E^i_b E^a_j -E^{ai}
E_{bj}) \left({\partial_i \log \Delta} \right)\, dy^j+{1\over 2} (\star_b d
\Delta^2)_{ijk}  E^{ai} E^j_b dy^k \cr && - {1\over 2}{\tau_2^{-1/2}}
{\rm Im} \left( \left[(F_{w_2} + \tau F_{w_1}) +i(H_{w_2} +\tau H_{w_1})
\right] E^{\bar w}  \right)_{ij} E^{ai}E^j_b  \cr {\Omega^w}_{-a} &= &~
{i\over 2 \tau_2} E^i_a
\partial_i \tau E^{\bar w}+{i \over 2 }\tau_2^{-1/2} \left[(F_{w_2} + \tau F_{w_1}) -i(H_{w_2}
+\tau H_{w_1}) \right]_{ij} E^i_a dy^j \cr {\Omega^w}_{-w}  &= &~ {i
\over 2 \tau_2} d \tau_1.
\eea

The aim is to solve the heterotic spinor equations. To do so we will
impose constraints on the fluxes $F_{w_i}$ and $H_{w_i}$ in such a
way that the spinor variations reduce to the ``flux-free''
situation. Note that all dependence on $F_{w_i}$ and $H_{w_i}$ will
cancel out if
\begin{equation} \label{cond2}
G_{\bar w}={1\over \sqrt{\tau_2}} {1\over 2 i} \left[ F_{w_2}  +
\tau F_{w_1} + i (H_{w_2} + \tau H_{w_1} )\right],
\end{equation}
and
\begin{equation} \label{cond3}
(G_w)^\star ={1\over \sqrt{\tau_2}} {1\over 2 i} \left[ F_{w_2}  +
\tau F_{w_1} - i (H_{w_2} + \tau H_{w_1} )\right],
\end{equation}
are primitive $(1,1)$  and $(2,0)$ forms on the base,
respectively. We recognize $G_w$ and $G_{\bar w}$ as the
components of the complex type IIB $3$-form $G_3$ expanded in an orthonormal frame
for the type IIB Calabi-Yau metric.

The conditions imposed on $G_w$ and $G_{\bar w}$ have a
natural interpretation as the conditions of imaginary self-duality
and primitivity of the complex $3$-form flux on the Calabi-Yau
space. Indeed, the requirement~\C{ISD}\ that the type IIB flux $G_3$
is imaginary self-dual implies
\begin{equation}\label{cond1}
F_{w_2}+\tau F_{w_1} + i \star_b (H_{w_2} + \tau H_{w_1})=0.
\end{equation}
Inserting this expression into the definition of  $G_w$ and $G_{\bar w}$
shows that
\begin{equation}
G_{\bar w}={1\over \sqrt{\tau_2}} {1\over 2 } \left[
(H_{w_2}-\star_b H_{w_2}) + \tau (H_{w_1}-\star_b H_{w_1}) )\right],
\end{equation}
is an anti-self-dual form on the base while
\begin{equation}
(G_w)^\star =-{1\over \sqrt{\tau_2}} {1\over 2 } \left[
(H_{w_2}+\star_b H_{w_2}) + \tau (H_{w_1}+\star_b H_{w_1}) )\right],
\end{equation}
is self-dual. Anti-self-dual $2$-forms on the base are primitive $(1,1)$
forms while self-dual forms can be of type $(2,0)$, $(0,2)$ and
non-primitive $(1,1)$. Requiring $G$ to be of Hodge type $(1,2)$\footnote{The reason we have a $(1,2)$ flux rather than
a $(2,1)$ flux as appears in~\cite{Dasgupta:1999ss}\ is the plus sign convention we have chosen for
$G_3$ given in eq.~\C{defineG}\ which is convenient for T-duality.} implies
that $G_w$ can only have a $(0,2)$ piece. These are precisely the two
conditions imposed on~\C{cond2}\ and~\C{cond3}\ to solve the spinor
variations on the heterotic side. The warp factor dependent term
appearing in the connection vanishes because the spinor has a definite
chirality on the base. Finally, the dilatino equation is solved by setting
\begin{equation}
e^{\hetdil}= \Delta.
\end{equation}

These are all the conditions needed to solve the $N=1$ space-time supersymmetry conditions. Note that
$\Delta$ is a scalar function of the base coordinates in the semi-flat approximation. Based on the supersymmetry
constraints alone, $\Delta$ is arbitrary. As we have
seen in section~\ref{tadpole}, the Bianchi identity gives a differential equation for $\Delta$ and its
solution determines the background completely.

\subsubsection{Semi-flat metric with two holomorphic parameters}

In this section, we will explore more general backgrounds of the
heterotic string. Heterotic string theory on $T^2$ has $18$ complex
moduli. One of these moduli is $\tau$ of the elliptic fiber while
$16$ correspond to the choice of Wilson lines for the
ten-dimensional gauge bundle. The remaining modulus, $\rho$, is the
(complexified) volume of the torus. If we fiber this modulus over
the base, we typically describe a non-geometric heterotic
compactification with torsion. A study of such solutions in relation
to F-theory will appear in~\cite{toappearjock}. These spaces are
intrinsically quantum since the patching conditions involve
T-duality rather than just diffeomorphisms. Nevertheless, we can
study whether such spaces are locally supersymmetric as in the
preceding sections.

There has been a reasonable study of non-geometric solutions. We will begin by
reviewing the metric of~\cite{Hellerman:2002ax}\ adapted to our notation. This is a generalization of the semi-flat
metric which depends on two holomorphic parameters. In the subsequent section, we will take
this solution as a starting point for constructing a large new class of
torsional backgrounds.

The metric simply has a varying volume for the torus fiber in addition to the $\tau$ fibration
\begin{equation}\label{dmetric}
ds^2 = g_{i j} dy^i dy^j +{\rho_2 \over \tau_2} \mid dw_2 + \tau
dw_1\mid^2,
\end{equation}
depending on two complex parameters $\tau=\tau_1 + i \tau_2$ and
$\rho= \rho_1+i \rho_2$, which are functions of the base coordinates
$y$ only. Moreover, there is ${\cal H}$-flux with two components in
the fiber directions
\begin{equation}
{\cal H}=d \rho_1\wedge  dw_2 \wedge dw_1,
\end{equation}
and the dilaton is related to $\rho$ according to
\begin{equation}
\hetdil = {1\over 2} \log \rho_2.
\end{equation}
We did not encounter torsion of this type in our preceding discussion.

As we will see below, supersymmetry requires the base to be K\"ahler.
Moreover, $\tau$ and $\rho$ have to be holomorphic functions of the
base coordinates. Let us solve the spinor equations. The spin
connections of the metric \C{dmetric} are
\begin{equation}
\begin{split}
& {\Omega^a}_b = {\omega^a}_b, \\ &{\Omega^w}_a = \half \partial_i
\log \rho_2 e^i_a E^w+{i\over 2} {\partial_i \tau \over \tau_2}
E^i_a E^{\bar w} ,\\ &  {\Omega^w}_w= {i \over 2} {d \tau_1 \over
\tau_2}.\\
\end{split}
\end{equation}
Using the expression for ${\cal H}$, the connection coefficients
arising in the gravitino supersymmetry variations are
\begin{equation}
\begin{split}
& {\Omega^a_-}_b={\omega^a}_b\cr & {\Omega^w_-}_a= {i\over 2}E^i_a
\left( {
\partial_i \rho\over  \rho_2}E^w +{\partial_i\tau \over \tau_2}
E^{\bar w}\right), \cr & {\Omega^w_- }_w = {i \over 2} \left(
{d\tau_1 \over \tau_2} +{d \rho_1 \over \rho_2} \right).
\end{split}
\end{equation}
The gravitino variation in the fiber directions takes the form
\begin{equation}
\begin{split}
& \delta \psi_{w_1} = \partial_{w_1} \xi + {i\over 4} \sqrt{\rho_2
\over \tau_2} \left(\tau {\partial_{\bar i}\rho \over \rho_2}+\bar
\tau {\partial_{\bar i}\tau  \over \tau_2}  \right) {\Gamma_w}^{\bar
i}\xi=0, \cr & \delta\psi_{w_2} = \partial_{w_2}\xi + {i\over 4}
\sqrt{\rho_2 \over \tau_2} \left({\partial_{\bar i} \rho \over
\rho_2} + { \partial_{\bar i}\tau  \over \tau_2} \right)
{\Gamma_w}^{\bar i}\xi=0.
\end{split}
\end{equation}
We will solve these conditions by requiring the spinor $\xi$ to be a
function of the base coordinates only, {\it i.e.}
\begin{equation}
\xi=\xi(y),
\end{equation}
and by imposing holomorphicity of $\rho$ and $\tau$
\begin{equation}
\bar \partial \rho=0 \qquad {\rm and } \qquad \bar\partial \tau=0.
\end{equation}
The spinor equations along the base are
\begin{equation}
\begin{split}\label{gravi}
& \delta \psi_i = \nabla_i \xi -{i\over 4}\left({\partial_i \tau_1
\over \tau_2} +{\partial_i \rho_1 \over \rho_2} \right) \xi=0,\cr &
\delta \psi_{\bar i} = \nabla_{\bar i} \xi -{i\over
4}\left({\partial_{\bar i} \tau_1 \over \tau_2} +{\partial_{\bar i}
\rho_1 \over \rho_2} \right) \xi=0.\cr
\end{split}
\end{equation}
This implies that the norm of $\xi$ is constant
\begin{equation}
\nabla_i (\xi^\dagger \xi)= \nabla_{\bar i} (\xi^\dagger \xi)=0,
\end{equation}
and therefore we can normalize $\xi^\dagger \xi=1$. Moreover, the
base is K\"ahler since $J_{mn} =-i \xi^\dagger {\Gamma_m}^n \xi$ is
covariantly constant and satisfies $J^2=-1$ (here $m,n$ are real
coordinates on the base). Using the holomorphicity of $\tau$ and
$\rho$, the integrability condition for a solution of~\C{gravi}\
is that the Ricci-form of the base is related to $\tau$ and $\rho$
according to
\begin{equation} \label{randone}
{\cal R}=i\partial \bar \partial \left( \log \tau_2+\log
\rho_2\right) ,
\end{equation}
or since the base is K\"ahler
\begin{equation} \label{randtwo}
\partial \bar \partial\left(  \det \log g_{i \bar j } - \log \tau_2-\log \rho_2\right) =0.
\end{equation}
The relation between the dilaton and $\rho$ can be derived taking
the dilatino variation into account.

\subsubsection{Torsional background with two holomorphic parameters}

We conclude our study of the local supersymmetry properties by generalizing the
solution depending on $\tau$ and $\rho$. Here we present the form of the background but
we will not repeat the supersymmetry analysis which closely follows the
previous three cases. However, we would like to emphasize that this
example illustrates that the heterotic background can be generalized in the sense of
``adding flux'' to the previously known solution.

In this case, the
flux is added to the solution with two holomorphic parameters.
The ${\cal H}$-flux which is added combines with the
modified spin connection arising from the twist of the
elliptic fiber in such a way that supersymmetry is still preserved.

The metric is
\begin{equation}
ds^2 = \Delta^2 g_{i j} dy^i dy^j +{\rho_2 \over \tau_2} \mid dw_2 +
\tau dw_1+A_H\mid^2,
\end{equation}
where
\begin{equation}
A_H=B_{w_2} + \tau B_{w_1}.
\end{equation}
The ${\cal H}$-flux is
\begin{equation}
{\cal H}=d \rho_1\wedge  dw_2 \wedge dw_1+{1\over\sqrt{
\tau_2\rho_2}} \, {\rm Im} \left[ (F_{w_2} + \tau F_{w_1} ) E^{\bar
w}\right] +\star_b d \Delta^2
\end{equation}
and the dilaton is
\begin{equation}
\hetdil = {1\over 2} \log\left( \rho_2 \Delta^2\right).
\end{equation}
This background is supersymmetric as long as~\C{randone}\ and~\C{randtwo}\
are satisfied, and $(H_{w_k}, F_{w_k})$ are subject
to the conditions described in section~\ref{toroneparameter}.

\section{Quantum Exact Metrics}
\label{desing}

In this final section, we turn to the issue of quantum corrected metrics and desingularization. The two issues are closely connected.  The semi-flat metrics we used to construct the torsional
solutions possess exact $U(1) \times U(1)$ isometries. For a compact Calabi-Yau metric, we know that there are no
such exact symmetries.
The breaking of these isometries can be viewed as resulting from quantum corrections to the
semi-flat metric, or equivalently, from gluing in smooth metrics to repair the singularities of the elliptic fibration.

The type IIB starting metric $\M$ is conformally Calabi-Yau. The conformal factor depends on the flux which
sets a physical length scale. If we examine the metric near near any degeneration of the elliptic fiber, there is a length
scale below which the flux is irrelevant. The singularities of the elliptic fibration of $\M$ can therefore always be repaired
by gluing in the
metrics employed in the case without flux with only minor modifications. There are no obstructions to smoothing the metric of the initial type IIB flux
compactification.

Nevertheless, little is actually known about the gluing metrics for elliptic spaces of complex dimension three. For complex dimension
two, there is an explicit metric found by Ooguri and Vafa~\cite{Ooguri:1996me}. The metric captures quantum corrections
that smooth a $\tau$ degeneration over ${\mathbb C}$ with coordinate $z$ where
\be \label{taudegeneration}
\tau(z) = {1\over 2\pi i} \log(z).
\ee
In the context studied, the quantum corrections can be viewed as arising from $D$-instantons.

The metric can also be obtained
directly and simply from gauge theory~\cite{Seiberg:1996ns}. The resulting quantum corrected
metric still possesses a single $U(1)$ isometry. For an elliptic $K3$ metric, gluing in this metric near each $\tau$ degeneration will locally preserve one particular $U(1)$ action but globally no $U(1)$ survives as we expect. This case has been examined in some detail in~\cite{gross-2000}.

It is natural to ask how quantum corrections modify elliptic metrics in the presence of torsion.
We should stress that these quantum corrections cannot be obtained by duality and are quite different on either side of the duality map.

In complex dimension two, there is an analogue of the Ooguri-Vafa
metric which we will now describe. The metric answers a roughly
T-dual version of the question of resolving a $\tau$ singularity of
the type given in~\C{taudegeneration}. The dual question is to
consider an elliptic metric for a space $\N$ of dimension two
through which we thread $N_f$ units of $H_3$-flux so that \be
\label{torsioncoulomb} \int_\N dH_3 = N_f. \ee This is the T-dual of
the metric charge measurable at infinity. The metric
singularity~\C{taudegeneration}\ defines a type IIB D7-brane. The
four-dimensional metric is the M-theory realization of the type IIB
D7-brane. Similarly, our metric describes $N_f$ NS5-branes localized
at points on $T^2 \times \RR^2$. Initially we will consider the case
where the branes are all coincident at the origin. Later we will add
the moduli for separating the branes.

Metrics of this kind appear explicitly in our constructions and the earlier ones of~\cite{Dasgupta:1999ss}\ where, for example, there is some $H_3$-flux through a $K3$ surface. At the level of supergravity, there is a net flux while the higher derivative correction to~\C{bianchi}\ eventually ensures that the total charge is zero. The local model for such a metric where we keep the elliptic fiber fixed but decompactify the base receives quantum corrections which we will now derive.

The quantum corrections are most naturally derived using gauge theory and we
will generalize a discussion of two-dimensional Coulomb branches given in~\cite{Diaconescu:1997gu}.
We consider a $U(1)$ $N=2$ gauge theory in four dimensions. On compactification on $T^2$, we obtain a $(4,4)$ gauge theory with moduli
space
\be
T^2 \times \RR^2.
\ee
The torus factor comes from the choice of Wilson line on $T^2$. We can view this $(4,4)$ theory as a special case of a $(2,2)$ theory.
The $(4,4)$ vector multiplet decomposes into a $(2,2)$ vector superfield whose field strength is a twisted chiral multiplet $\Sigma$,
together with a chiral multiplet $\Phi$. These complex fields are coordinates for $T^2 \times \RR^2$ with the compact directions captured by $\Sigma$.

Non-linear sigma models with $(2,2)$ supersymmetry are characterized
by a generalized K\"ahler potential $K(\Sigma, \Phi)$; see, for
example,~\cite{Gates:1984nk}. The potential satisfies
\be\label{analma} K_{\Phi {\bar \Phi}} + K_{\Sigma \bar\Sigma} = 0
\ee with metric \be ds^2 = K_{\Phi {\bar \Phi}} |d\Phi |^2 -
K_{\Sigma \bar\Sigma} | d\Sigma |^2 \ee and torsion \be {\cal B} =
{1\over 4} \left( K_{\Phi \bar\Sigma} d\Phi d\bar\Sigma +
K_{\Sigma\bar\Phi} d\bar\Phi d\Sigma \right). \ee The
constraint~\C{analma}\ is the analogue of the usual Monge-Amp\`ere
condition for Calabi-Yau spaces. The classical flat generalized
K\"ahler potential is just \be K_0 = {1\over g^2} \left( \Phi
\bar\Phi - \Sigma\bar\Sigma \right) \ee where $g$ is the
two-dimensional gauge coupling. So far we have just reproduced well
known facts nicely summarized in~\cite{Diaconescu:1997gu}. It is
quite beautiful, though, that gauge theory in two dimensions
naturally gives torsional metrics.

Now we would like to examine quantum corrections. Let us couple $N_f$ charged hypermultiplets to the abelian gauge theory. This will lead to torsion on the Coulomb branch which is measured by~\C{torsioncoulomb}.

Viewed as a $(2,2)$ multiplet, each hypermultiplet contains two chiral superfields $(q_1, q_2)$.  The metric on the Coulomb branch is $1$-loop exact so all
quantum corrections are captured by integrating out the hypermultiplets. Since we are considering compactified gauge theory, we must perform a sum over the Kaluza-Klein modes of the hypermultiplet on the $T^2$.

For simplicity, let us take the world-volume for the gauge theory to be $T^2 = S^1_{R_1} \times S^1_{R_2}$ rectangular. The $1$-loop correction to the gauge kinetic term
is given by
\be\label{kksum}
N_f \sum_{n_1, n_2 = -\infty}^{+\infty} \, \int {d^2 k \over (2\pi)^2 R_1 R_2} \, {1\over \left( k^2  + |\Phi |^2 + ({n_1\over R_1} + \sigma_1)^2 + ({n_2\over R_2} + \sigma_2)^2 \right)^2}
\ee
where $\Sigma = \sigma_1 + i \sigma_2$. The periodicities of the compact coordinates are given as follows: $\sigma_1 \sim \sigma_1 + {1\over R_1}$ and $\sigma_2 \sim \sigma_2 + {1\over R_2}$.

Applying Poisson resummation to~\C{kksum}\ gives the $1$-loop corrected gauge coupling:
\bea\label{gaugecoupling}
 && K_{\Sigma\bar\Sigma}= {1\over g^2} +   R_1 R_2  { N_f \over 2}  \times   \bigg[  \log \left( {\Lambda \over |\Phi |} \right)  \, +   \cr   && \sum_{m_1,m_2 \neq 0} K_0(2\pi \omega |\Phi|) \times e^{2\pi i \left( m_1\sigma_1 R_1 + m_2 \sigma_2 R_2 \right)} \bigg]
\eea
where $\omega = \sqrt{ (m_1 R_1)^2 + (m_2 R_2)^2}$ and $\Lambda$ is some constant scale. The gauge coupling is given by $K_{\Sigma\bar\Sigma}$ and so~\C{gaugecoupling} determines the quantum corrected metric and $K$.

The modified Bessel function $K_0(z) \sim \sqrt{\pi \over 2 z} e^{-z}$ for large $z$ so the quantum corrections look precisely like instanton corrections to the metric. Unlike the case of the $\tau$ degeneration, the presence of torsion breaks both isometries of the fiber.

It is interesting that we also find $K_0(z)$ capturing the quantum
corrections like the case studied in~\cite{Ooguri:1996me}. However,
the reason is very different. In that case, from the gauge theory
perspective one performs an integral over three-dimensional loop
momenta and then a one-dimensional Poisson resummation. In our case,
we perform an integral over two-dimensional loop momenta and then a
two-dimensional Poisson resummation.

Finally we can introduce complex masses $\Phi_i$ for the $N_f$ hypermultiplets and Wilson line moduli $(\sigma_1^i, \sigma_2^i)$. These moduli correspond to splitting the locations of the $N_f$ branes.  The resulting metric is the sum of $1$-loop corrections localized around each brane
\bea
\label{finalmetric}
 && K_{\Sigma\bar\Sigma} = {1\over g^2} +  { R_1 R_2 \over 2}  \times \sum_{i=1}^{N_f}   \bigg[  \log \left( {\Lambda \over |\Phi - \Phi_i |} \right)  \, +   \cr   && \sum_{m_1,m_2 \neq 0} K_0(2\pi \omega |\Phi - \Phi_i|) \times e^{2\pi i \left( m_1 \left( \sigma_1 - \sigma_1^i \right) R_1 + m_2 \left( \sigma_2 - \sigma_2^i \right) R_2 \right)} \bigg].
\eea The key difference between the torsional and non-torsional
cases is the breaking of both $U(1)\times U(1)$ isometries for
NS-branes versus the breaking of just one isometry for the smoothed
D7-brane metric of~\cite{Ooguri:1996me}. A similar breaking of a
$U(1)$ isometry appears in the relation between localized NS5-branes
on a circle and KK monopoles~\cite{Gregory:1997te, Tong:2002rq}.

Now the analysis for complex dimension three metrics is much more challenging. In that case of prime interest, a theorem is needed to prove the existence of smooth metrics of the kind described in~\C{wwwx}. That would be evidence that the non-linear sigma model on such a space flows to a superconformal field theory. It is also essential to find a tractable world-sheet description of those models which would permit the computation of correlators. Perhaps along the lines examined in~\cite{Adams:2006kb, Adams:2009av, toappearsav}.


\section*{Acknowledgements}
It is our pleasure to thank Aaron Bergman, Guangyu Guo, Chris Hull, Jock McOrist, Greg Moore, David R. Morrison, Callum Quigley and Ashoke Sen for helpful discussions. We are particularly grateful to Oleg Lunin for comments and discussions. We would also like to thank the Aspen Center for Physics for hospitality during the completion of this project.  K.~B. would like to thank the KITP for hospitality during the final stages of this work.

\vskip 0.1 in
\noindent
The work of K. B. is supported in part by NSF Grant No. PHY-0505757 and by NSF Grant No. PHY-05-51164. The work S.~S. is supported in part by NSF Grant No. PHY-0758029  and by NSF Grant No.~0529954.

\newpage
\appendix
\label{appendix}

\section{Notation}
\label{notation}

\begin{itemize}
\item

We use
$$M,N,
\dots \mu,\nu,\dots, i,j,\dots, w_1,w_2,\quad (A,B,\dots
\alpha,\beta,\dots, a,b,\dots , w ,\bar w)$$ to denote the
coordinate bases of any six-dimensional space, of four-dimensional Minkowski space-time, and separately of the base and fiber of an elliptic six-dimensional space,
respectively.

For coordinates on the four-dimensional base of a six-dimensional elliptic space, we use $y^i$ while we denote the fiber coordinates by $w_i$, $i=1,2$.

The indices included in parentheses are with respect to
an orthonormal rather than coordinate basis.

\item
We will use ${\cal H}$ or
${\cal H}_3$ to denote the heterotic NS flux and $H$ or $H_3$ to denote the
type II NS flux. The associated gauge potentials are denoted ${\cal
B}_2$ and $B_2$, respectively. The standard notation $F_n$ will be used for the RR
fluxes of type II string theory defined in~\C{deffieldstrengths}\
with associated potentials $C_{n-1}$. For type I string theory,  we
use the notation $F_n'$ for the RR fluxes.

\item In the supersymmetry transformations we use the notation
$$
 /\!\!\! \!{\cal
H}_{M}= {1\over 2} {\cal H}_{MNP} \Gamma^{NP}, \qquad
/\!\!\! \!{ \cal
H}={1\over 3!} {\cal H}_{MNP} \Gamma^{MNP}.
$$

\item
To describe the various fluxes we use the index notation
\bea
F_{w_k}& = &{1\over 2!} F_{y_i y_j w_k} dy^i \wedge dy^j,  \qquad k=1,2\cr
H_{w_k}& = &{1\over 2!} H_{y_i y_j w_k} dy^i \wedge dy^j,  \qquad k=1,2\cr
C_{w_k}& = & C_{y_i w_k}  dy^i, \qquad k=1,2\cr
B_{w_k}& = & B_{y_i w_k}  dy^i, \qquad k=1,2.\cr
\non
\eea

\end{itemize}

\section{T-duality Rules}

The Buscher rules for T-dualizing in the $x$ direction are given by:
\bea
e^{2\Phi'} &=& \frac{e^{2\Phi}}{G_{xx}}, \non\\
G_{xx}' &=& {1\over G_{xx}}, \non\\
G_{M x}' &=& {B_{M x}\over G_{xx}}, \\
G_{MN}' &=& G_{MN} - \frac{G_{M x}G_{N x} - B_{M x}B_{N x}}{G_{xx}},
\non\\
B_{M x}' &=& {G_{M x}\over G_{xx}}, \non\\
B_{MN}' &=& B_{MN} - \frac{B_{M x}G_{N x} - G_{M x}B_{N x}}{G_{xx}}.
\non
\eea
The transformation of the R-R potentials is given by:
\bea
{C_{M\cdots N P  x}^{(n)}}' &=& C_{M\cdots N P }^{(n-1)} -
(n-1)\frac{C_{[M\cdots N|x}^{(n-1)}G_{| P ]x}}{G_{xx}}, \\
 {C_{M\cdots N P  Q}^{(n)}}' &=&
C_{M\cdots N P  Q x}^{(n+1)} + n{C_{[M\cdots N P |x| }^{(n)}}'B_{ Q]x}. \non\\
\eea
For convenience we present the Buscher rules for the field strengths $F^{(n)}$ which, in supergravity,
are related to the potentials as follows
\be\label{deffieldstrengths}
F^{(n)} = dC^{(n-1)} + H \wedge C^{(n-3)},
\ee
where $H=dB$ is the NS-NS field strength.
Using the results for the T-dual potentials we find
\bea
{F^{(n)}_{M\dots NP x} }'& =& {F^{(n-1)}_{M\dots NP } } -(n-1)\frac{F_{[M\cdots N|x}^{(n-1)}G_{| P ]x}}{G_{xx}}, \non\\
{F^{(n)}_{M \dots NPQ}}'& =& {F^{(n+1)}_{M \dots NPQx}}+ n{F_{[M\cdots N P|x| }^{(n)}}'B_{ Q]x}.
\eea

\newpage


\providecommand{\href}[2]{#2}\begingroup\raggedright\endgroup

\end{document}